%% file: sample-journal.tex
\documentclass[manuscript,screen]{acmart}
\usepackage{booktabs} 

\usepackage[ruled]{algorithm2e} 
\usepackage{multirow}

\usepackage{tikz}
\usetikzlibrary{shapes,arrows,chains}
\usepackage{pgfplots}

\usepackage{pdfcomment}

\usepackage{tablefootnote}

\begin{document}
\title{Unrolling Ternary Neural Networks}

\author{Stephen Tridgell}
\affiliation{%
  \institution{The University Of Sydney}
  \streetaddress{School of Electrical and Information Engineering, Building J03}
  \city{The University Of Sydney}
  \state{NSW}
  \postcode{2006}
  \country{Australia}}
\email{stephen.tridgell@sydney.edu.au}

\author{Martin Kumm}
\affiliation{%
  \institution{Fulda University Of Applied Sciences}
  \country{Germany}}
\email{martin.kumm@cs.hs-fulda.de}

\author{Martin Hardieck}
\affiliation{%
  \institution{University Of Kassel}
  \country{Germany}}
\email{hardieck@uni-kassel.de}

\author{David Boland}
\affiliation{%
  \institution{The University Of Sydney}
  \streetaddress{School of Electrical and Information Engineering, Building J03}
  \city{The University Of Sydney}
  \state{NSW}
  \postcode{2006}
  \country{Australia}}
\email{david.boland@sydney.edu.au}

\author{Duncan Moss}
\affiliation{%
  \institution{The University Of Sydney}
  \streetaddress{School of Electrical and Information Engineering, Building J03}
  \city{The University Of Sydney}
  \state{NSW}
  \postcode{2006}
  \country{Australia}}
\email{dunncan.moss@sydney.edu.au}

\author{Peter Zipf}
\affiliation{%
  \institution{University Of Kassel}
  \country{Germany}}
\email{zipf@uni-kassel.de}

\author{Philip H.W. Leong}
\affiliation{%
  \institution{The University Of Sydney}
  \streetaddress{School of Electrical and Information Engineering, Building J03}
  \city{The University Of Sydney}
  \state{NSW}
  \postcode{2006}
  \country{Australia}}
\email{philip.leong@sydney.edu.au}

\renewcommand\shortauthors{Tridgell et al}

\begin{abstract}
The computational complexity of neural networks for large scale or real-time applications necessitates hardware acceleration.
Most approaches assume that the network architecture and parameters are unknown at design time, permitting usage in a large number of applications.
This paper demonstrates, for the case where the neural network architecture and ternary weight values are known a priori, that extremely high throughput implementations of neural network inference can be made by customising the datapath and routing to remove unnecessary computations and data movement.
This approach is ideally suited to FPGA implementations as a specialized implementation of a trained network improves efficiency while still retaining generality with the reconfigurability of an FPGA.
A VGG style network with ternary weights and fixed point activations is implemented for the CIFAR10 dataset on Amazon's AWS F1 instance.
This paper demonstrates how to remove 90\% of the operations in convolutional layers by exploiting sparsity and compile-time optimizations.
The implementation in hardware achieves $90.9 \pm 0.1$\% accuracy and 122\,k frames per second, with a latency of only 29\,\textmu s, which is the fastest CNN inference implementation reported so far on an FPGA.
\end{abstract}

\setcopyright{acmlicensed}
\acmJournal{TRETS}
\acmYear{2018} \acmVolume{1} \acmNumber{1} \acmArticle{1} \acmMonth{1} \acmPrice{15.00}
\acmDOI{0000001.0000001}



\maketitle

\input{samplebody-journals}

\end{document}

%% file: samplebody-journals.tex
\section{Introduction}
The computational complexity of convolutional neural networks (CNN) imposes limits to certain applications in practice \cite{jouppi2017datacenter}.
There are many approaches to this problem with a common strategy for the inference problem being to reduce the precision of arithmetic operations, or to increase sparsity \cite{mellempudi2017ternary,
faraone2017compressing,
boo2017structured,
courbariaux2016binarized,
rastegari2016xnor,
zhou2016dorefa}.
It has been shown that low precision networks can achieve comparable performance to their full precision counterparts \cite{li2016ternary,
courbariaux2016binarized,
zhou2016dorefa}.

This paper explores further accelerating inference with Field Programmable Gate Array (FPGA) implementations of neural networks in fixed point arithmetic with ternary weight values.
As the weights are restricted to $\{-1,0,1\}$, the multiplications between the inputs and the weights are reduced in complexity to subtractions, additions or no operation.
Compared with previous work, the datapath is customised based on known weight values of a trained network at design time.

The computationally intensive parts of a CNN are matrix-vector multiplications.
Given prior knowledge of the ternary weight matrix, this paper shows matrix vector multiplications can be implemented as the evaluation of an unrolled and pruned adder tree, effectively storing the weights in the routing.
This technique is only recently feasible due to increasing sizes of FPGAs, improvements in the tools and advances in low precision machine learning.
The main contributions of this paper are as follows:
\begin{itemize}
\item A novel architecture for inference of ternary neural networks on an FPGA by employing complexity reduction to the evaluation of pruned adder trees.
In particular optimisations are: (1) streaming inputs and full pipelining of the computation; (2) ternary weights with 16-bit sums and activations to preserve accuracy; (3) bit-serial summation for compactness; (4) weight-specific circuit generation with removal of multiplies by 0 and merging of common subexpressions to minimise computation; (5) throughput matching of CNN layers to minimise resource usage.
\item An extension of a technique to train ternary neural networks described by Li et al. \cite{li2016ternary} allowing simple control of the sparsity of the network.
\item A publicly available code with the highest reported throughput and lowest latency over the best published results on the CIFAR10 \cite{krizhevsky2009learning} benchmark, while achieving superior accuracy over other low-precision CNNs.
\end{itemize}

\section{Background}

\subsection{Deep Neural Networks \& Convolutions}
Deep Neural Networks (DNNs)~\cite{DL} are a class of machine learning algorithms that are described as a connected graph of basic compute nodes called neurons.
Each layer $l$ performs a mapping $\mathbb{R}^M \rightarrow \mathbb{R}^N$, where $\mathbf{x^l}$ is the input vector, $\mathbf{a^l}$ is the output or activation vector and $b_{i}^{l}$ is the bias term.
The computation for each neuron is the inner product between the input and its weight vector, $\mathbf{w_{i}^l} \in \mathbb{R}^N$, followed by an activation function $f$ which is typically tanh, ReLU, sigmoid, etc. 
The operation performed to compute the $i$'th output in layer $l$ is computed as
\begin{equation}
a_{i}^{l} = f(\mathbf{w_{i}^{l}} \mathbf{x^{l}} + b_{i}^{l}) 
\label{equ:nn_neuron}
\end{equation}
DNNs are arranged such that the output from one layer is passed to the input of neurons in a subsequent layer, these are named dense layers. 
By stacking multiple dense layers together, complex nonlinear interactions can be modeled; these network topologies are named multilayer perceptrons (MLP).
If inference is performed with a single input vector at a time, the computation for each layer can be described as a matrix-vector multiplication.
Consequently, the main bulk of the computation in an MLP requires successively performing matrix-vector multiplications with the layers' weights $w^{l}$ and the previous layer's activations $\mathbf{a^{l-1}}$.

Convolutional Neural Networks (CNNs) provide a way of reducing the number of parameters in the network while also exploiting the spatial property in images where pixels located close together are more related.
Let $x$ be a square input image with height and width $W$ and a depth of $D$.
For example, an input image might have $32 \times 32$ pixels where each pixel has a Red, Green, Blue (RGB) component, hence having $D=3$.
The convolution operation partitions the image into small overlapping sections of $N \times N$ pixels where typically $N \ll W$.
On each of these sections, $F$ different filters are applied and the results stacked together giving an output image of depth $F$.
This can be formalised as given an input image $x \in \mathbb{R}^{W\times W\times D}$, the convolutional weights $w \in \mathbb{R}^{N\times N\times D\times F}$ apply a transformation to give an output image $y \in \mathbb{R}^{W\times W\times F}$.
The calculation of the convolution operation is given as
\begin{equation}
y_{i,j,f} = \sum_{q=0}^{N} \sum_{r=0}^{N} \sum_{s=0}^{D} x_{a,b,s} w_{q,r,s,f} 
\label{equ:cnn_conv}
\end{equation}
where $a = i+q-\lfloor N/2 \rfloor$ and $b = j+r-\lfloor N/2 \rfloor$ and assuming the image is padded with zeros around the borders.
Expanding the sums in Equation \ref{equ:cnn_conv} the input $x$ can be transformed into a vector of length $N \times N \times D$ for a chosen $i$ and $j$ denoted $\mathbf{X}_{i,j}$.
The weights are independent of $i$ and $j$ and are rearranged as a matrix of constants with $N \times N \times D$ columns and $F$ rows to produce the output vector $\mathbf{y}_{i,j}$ of length $F$.
This can be written as
\begin{equation}
\mathbf{y}_{i,j} = \mathbf{X}_{i,j} \mathbf{w}
\label{equ:cnn_conv_unrolled}
\end{equation}
which is a matrix-vector operation for each pixel of the image.
An important property of the convolution that is exploited in this paper is that the result for the convolution only depends on a small area around the coordinates $i,j$ and that the weights, $\mathbf{w}$, are constant in regard to the chosen coordinates.
A hardware block can now be designed to take $\mathbf{X}_{i,j}$ as input and output $\mathbf{y}_{i,j}$ as a parallel pipeline.

\subsection{Low Precision Networks}
Interest in low precision CNNs has dramatically increased in recent years due to research which has shown that similar accuracy to floating point can be achieved \cite{mellempudi2017ternary,
faraone2017compressing,
boo2017structured,
courbariaux2016binarized,
rastegari2016xnor,
zhou2016dorefa}. 
Due to the high computational requirements of CNNs, reduced precision implementations offer opportunities to reduce hardware costs and training times.
Since FPGAs can implement arbitrary precision datapaths, they have some advantages over the byte addressable GPUs and CPUs for these applications.
Moreover, the highest throughput implementations on all platforms utilise reduced precision for a more efficient implementation.
In ternary neural networks, implemented similarly to Li et al. \cite{li2016ternary}, the value of the weights is restricted to be $-s$, 0 or $s$, where $s$ is a scaling factor.
This transforms Equation \ref{equ:nn_neuron} to
\begin{equation}
a_i = f(s^{l} \mathbf{w_{i}^{l}} \mathbf{x^{l}} + b_{i}^{l})
\label{equ:tnn_neuron}
\end{equation}
where $w_{ij}^{l} \in \{-1, 0 ,1\}$ are the elements of $\mathbf{w_{i}}$, $b_{i}^{l}$ is the bias term and $s^{l} \in \mathbb{R}$.

Applying it to the convolution operation in Equation \ref{equ:cnn_conv_unrolled}, it can be rewritten with ternary weights $\mathbf{t}$ and scaling factor $s$ as
\begin{equation}
\mathbf{y}_{i,j} = s \mathbf{X}_{i,j} \mathbf{t}
\label{equ:tnn_conv_unrolled}
\end{equation}
The matrix of values for the ternary weights, $\mathbf{t}$, are constant, restricted to $-1, 0, 1$ and known at design time.
The input $\mathbf{X}_{i,j}$ is an unknown dense vector and $s$ is a constant scalar as defined above.
Therefore a hardware block can be designed to take $\mathbf{X}_{i,j}$ as input and output $\mathbf{y}_{i,j}$.
With these assumptions, this paper demonstrates that this hardware block can be implemented with parallel adder trees to skip a large portion of the computation.
The consideration of the ternary convolutional operation in this manner is a core contribution of this paper and the benefits of this approach particularly in regard to unstructured sparsity are demonstrated in subsequent sections.

\subsection{Hardware Accelerators of CNNs}

Previous work \cite{
jouppi2017datacenter,
chen2017eyeriss,
Wang2017ReconfigurablePF,
qiu2016going,
baskin2018streaming,
meloni2018neura,
zhang2017frequency,
moss2017high,
li20177,
venkatesh2017accelerating,
umuroglu2017finn,
fraser2017scaling,
liang2018fp,
kimfpga,
prost2017scalable} has mainly focused on general accelerators, implemented either as a sequence of instructions on fixed hardware, or accelerator platforms designed for linear algebra intensive computation.
Performance comparisons with those in this paper are presented in Section~\ref{se:results}.

Systolic array based architectures implement a grid of local-connected processing units to perform a matrix-to-matrix multiplication.
Most notably, Jouppi et al. \cite{jouppi2017datacenter} describe the Tensor Processing Unit (TPU), an Application Specific Integrated Circuit (ASIC) that utilises a systolic array to compute operations necessary with 8 or 16 bit weights and activations.
It provides a very high throughput on a variety of applications with a latency on the scale of 10\,ms, claiming a peak throughput of 92\,TOps/sec and a sustained throughput of 86\,TOps/sec.
Similarly, Moss et al. \cite{moss2017high} implemented a systolic array that utilised the Intel QuickAssist heterogeneous system with a CPU and FPGA, accelerating CNNs with binary activations and weights on ImageNet and achieving 40.96\,TOps/sec.
Venkatesh et al. \cite{venkatesh2017accelerating} use a method described in \cite{li2016ternary} to implement a VGG style network with ternary weights and half-precision floating point activations.
They create an ASIC accelerator for training networks in addition to inference with a systolic array like structure.
Due to the use of floating point activations, they achieve high accuracy on CIFAR10 of around 91\%.

Differing from the systolic array approach, vector processors contain several independent processing lanes, with the capability of each determined by the lanes' architecture.
Chen et al. \cite{chen2017eyeriss} created a custom ASIC utilising 16 bit fixed point achieving a peak throughput of up to 42\,GMAC/s. 
Their architecture contains an array of independent processing elements, each receiving operation instructions.
In the programmable logic domain, Wang et al. \cite{Wang2017ReconfigurablePF}, Qiu et al. \cite{qiu2016going} and Meloni et al. \cite{meloni2018neura} presented neural network accelerators for the Zynq CPU+FPGA.
Each implemented a vector processor and utilised low precision to improve computational performance for object detection and image recognition.
Wang et al. \cite{Wang2017ReconfigurablePF}, Qiu et al. \cite{qiu2016going} and Meloni et al. \cite{meloni2018neura} achieved 2\,TOps/sec, 187\,GOps/sec and 169\,GOps/sec respectively.
Finally, Zhang and Prasanna \cite{zhang2017frequency} implemented an accelerator using the frequency domain representation of the convolution.
Their datapath converts the activations into the frequency domain and uses a point-wise multiplication to perform the convolution; this is followed by an inverse FFT.
The computation is performed in floating point and implemented on the Intel QuickAssist platform containing a CPU and Stratix V FPGA, achieving 123.5\,GFLOPs.



\section{Architecture}

This section describes how different components of a CNN can be implemented on an FPGA.
For this paper the design is implemented similarly to Prost-Boucle et al. \cite{prost2017scalable} where the image is streamed through the network using dedicated blocks for each section of the network.
Under this assumption a CNN can be divided into a few logical blocks:
\begin{itemize}
\item Buffering - (im2col/im2row)
\item Max Pool
\item Convolution
\item Scale and Shift - (Batch Normalization Inference)
\item MUX Layers
\item Dense Layers
\end{itemize}
These blocks are described in the following sections.

\subsection{Buffering}
\label{sec:buffering}
There are two main approaches to compute the convolution.
They are either passing in inputs over multiple cycles and storing intermediate results when computing the convolution, or buffering the pixels so that the entire set of inputs needed for the convolution is available simultaneously such as the approach in Baskin et al. \cite{baskin2018streaming} and Prost-Boucle et al. \cite{prost2017scalable}.
This paper adopts the scheme used by these authors to buffer the pixels, so the entire set of inputs is available simultaneously.
It buffers previous inputs in such a way that each cycle it can output the current vector $\mathbf{X}_{i,j}$ from Equation \ref{equ:tnn_conv_unrolled}.
The inputs necessary to compute an output pixel of the convolution are only a small segment of the image.
As the image is streamed into a layer of the CNN it is buffered in order to transform the pixels to a patch of the image from which a convolution or Max Pool operation can be computed.
This is referred to as a buffering block in this paper but is also known as the im2row or im2col transform.
The purpose of the buffering hardware block is to stream the image through a fully pipelined design with $p$ pixels provided each cycle as input from a FIFO.
To compute an output pixel of the convolution requires using inputs from various pixels of the image.

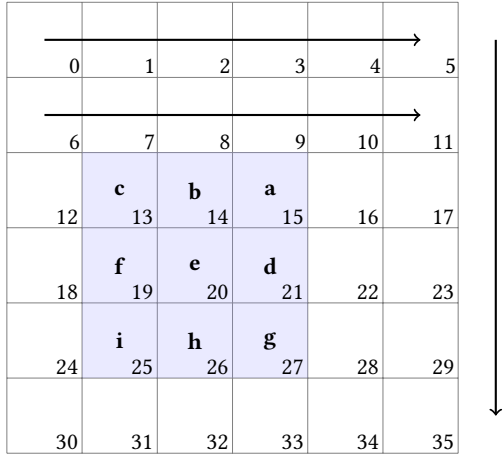
\begin{figure}
\centering
\begin{tikzpicture}
\draw[step=1cm,gray,very thin] (0,0) grid (6,6);
\fill[blue!40!white, fill opacity=0.2] (1,1) rectangle (4,4);
\draw[thick,->] (0.5,5.5) -- (5.5,5.5);
\draw[thick,->] (0.5,4.5) -- (5.5,4.5);
\draw[thick,->] (6.5,5.5) -- (6.5,0.5);
\node at (3.5,3.5) {\textbf{a}};
\node at (2.5,3.5) {\textbf{b}};
\node at (1.5,3.5) {\textbf{c}};
\node at (3.5,2.5) {\textbf{d}};
\node at (2.5,2.5) {\textbf{e}};
\node at (1.5,2.5) {\textbf{f}};
\node at (3.5,1.5) {\textbf{g}};
\node at (2.5,1.5) {\textbf{h}};
\node at (1.5,1.5) {\textbf{i}};

\node at (0.9,5.15) {0};
\node at (1.9,5.15) {1};
\node at (2.9,5.15) {2};
\node at (3.9,5.15) {3};
\node at (4.9,5.15) {4};
\node at (5.9,5.15) {5};

\node at (0.9,4.15) {6};
\node at (1.9,4.15) {7};
\node at (2.9,4.15) {8};
\node at (3.9,4.15) {9};
\node at (4.8,4.15) {10};
\node at (5.8,4.15) {11};

\node at (0.8,3.15) {12};
\node at (1.8,3.15) {13};
\node at (2.8,3.15) {14};
\node at (3.8,3.15) {15};
\node at (4.8,3.15) {16};
\node at (5.8,3.15) {17};

\node at (0.8,2.15) {18};
\node at (1.8,2.15) {19};
\node at (2.8,2.15) {20};
\node at (3.8,2.15) {21};
\node at (4.8,2.15) {22};
\node at (5.8,2.15) {23};

\node at (0.8,1.15) {24};
\node at (1.8,1.15) {25};
\node at (2.8,1.15) {26};
\node at (3.8,1.15) {27};
\node at (4.8,1.15) {28};
\node at (5.8,1.15) {29};

\node at (0.8,0.15) {30};
\node at (1.8,0.15) {31};
\node at (2.8,0.15) {32};
\node at (3.8,0.15) {33};
\node at (4.8,0.15) {34};
\node at (5.8,0.15) {35};

\end{tikzpicture}
\caption{Streaming the Image and the Window to Buffer}
\label{fig:imageWindow}
\end{figure}

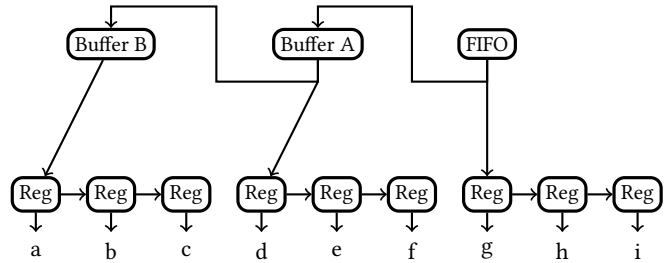
\begin{figure}
\centering
\begin{tikzpicture}

\tikzstyle{mybox} = [draw=black, very thick,
    rectangle, rounded corners, scale=0.9]

\node [mybox] (FIFO) { FIFO };
\node [mybox] at (-2.25, 0) (bufferA) { Buffer A };
\node [mybox] at (-5, 0) (bufferB) { Buffer B };

\draw[->,thick] (FIFO) -- ( 0,-0.5 ) -- ( -1, -0.5 ) -- ( -1, 0.5 ) -- ( -2.25, 0.5 ) -- (bufferA);
\draw[->,thick] (bufferA) -- ( -2.25,-0.5 ) -- ( -3.6, -0.5 ) -- ( -3.6, 0.5 ) -- ( -5, 0.5 ) -- (bufferB);

\node [mybox] at (2, -2) (srF0) {Reg};
\node [mybox] at (1, -2) (srF1) {Reg};
\node [mybox] at (0, -2) (srF2) {Reg};
\node [mybox] at (-1, -2) (srA0) {Reg};
\node [mybox] at (-2, -2) (srA1) {Reg};
\node [mybox] at (-3, -2) (srA2) {Reg};
\node [mybox] at (-4, -2) (srB0) {Reg};
\node [mybox] at (-5, -2) (srB1) {Reg};
\node [mybox] at (-6, -2) (srB2) {Reg};

\draw[->,thick] (FIFO) -- ( 0,-0.5 ) -- (srF2);
\draw[->,thick] (srF2) -- (srF1);
\draw[->,thick] (srF1) -- (srF0);
\draw[->,thick] (srF2) -- (0, -2.5);
\draw[->,thick] (srF1) -- (1, -2.5);
\draw[->,thick] (srF0) -- (2, -2.5);

\draw[->,thick] (bufferA) -- ( -2.25,-0.5 ) -- (srA2);
\draw[->,thick] (srA2) -- (srA1);
\draw[->,thick] (srA1) -- (srA0);
\draw[->,thick] (srA2) -- (-3, -2.5);
\draw[->,thick] (srA1) -- (-2, -2.5);
\draw[->,thick] (srA0) -- (-1, -2.5);

\draw[->,thick] (bufferB) -- (srB2);
\draw[->,thick] (srB2) -- (srB1);
\draw[->,thick] (srB1) -- (srB0);
\draw[->,thick] (srB2) -- (-6, -2.5);
\draw[->,thick] (srB1) -- (-5, -2.5);
\draw[->,thick] (srB0) -- (-4, -2.5);

\node at (-6, -2.75 ) (a) {a};
\node at (-5, -2.75 ) (b) {b};
\node at (-4, -2.75 ) (c) {c};
\node at (-3, -2.75 ) (d) {d};
\node at (-2, -2.75 ) (e) {e};
\node at (-1, -2.75 ) (f) {f};
\node at (0, -2.75 ) (g) {g};
\node at (1, -2.75 ) (h) {h};
\node at (2, -2.75 ) (i) {i};

\end{tikzpicture}
\caption{Diagram illustrating buffering for a $3 \times 3$ convolution}
\label{fig:convBuffer}
\end{figure}

Figure \ref{fig:imageWindow} demonstrates the input configuration used in this paper with the assumptions that the convolutional kernel is $3 \times 3$, the image is $6 \times 6 \times 1$ and a single pixel is streamed in each cycle.
The image is streamed in left to right, top to bottom with a pixel arriving in the cycle indicated by the number in each box.
The input \textit{FIFO} in Figure \ref{fig:convBuffer} outputs the pixel each cycle to both \textit{Buffer A} and the first stage of a shift register.
The value is then shifted across each cycle by the registers as the window moves across the image.
\textit{Buffer A} and \textit{Buffer B} delay the output by the image width (in the case of Figure \ref{fig:imageWindow} this is 6 cycles) in order to output the previous two rows of the image with matching columns.
For example, when the \textit{FIFO} outputs pixel $27$ in Figure \ref{fig:imageWindow}, \textit{Buffer A} outputs pixel $21$ and \textit{Buffer B} outputs pixel $15$.
After the outputs $a$-$i$ are obtained at the bottom of Figure \ref{fig:convBuffer}, there is additional logic necessary to implement zero padding by switching between these values and zero.
This produces the correct vector $\mathbf{X}_{i,j}$ from Equation \ref{equ:tnn_conv_unrolled} for the convolution.
This can also be adapted for \textit{Max Pool} layers where the structure of Figure \ref{fig:convBuffer} changes depending on the kernel size, image size and pixel rate $p$.

\subsection{Max Pool}
Max Pool layers are widely used in CNNs to downsample the image. 
The Max Pool operation takes a $k \times k$ window of the image as input similar to that shown in Figure \ref{fig:imageWindow}.
For these pixels, it compares them based on each component and outputs the maximum value resulting in a single pixel.
Reduction in the image size is achieved by using a stride greater than $n=1$.
Consider an example based on the image in Figure \ref{fig:imageWindow} with a kernel size of $k=2$.
If the stride is only $n=1$, the window of pixels $0,1,6,7$ would be followed by pixels $1,2,7,8$ meaning the output image is about the same size.
To downsample the image it is common to use a stride of $n=2$ or more in Max Pool layers.
With a stride of $n=2$, the window of pixels $0,1,6,7$ is followed by pixels $2,3,8,9$ which would produce a $3 \times 3$ image from the input in Figure \ref{fig:imageWindow}.
The presence of Max Pool layers has the effect of reducing the amount of computation required for subsequent layers of the CNN.
The downsampling property of Max Pool is exploited in our proposed architecture to reduce hardware area.
Given a throughput of $p$ pixels/cycle as input into a Max Pool layer with a stride of $n$ results in the output throughput reduced to $\frac{p}{n^2}$ pixels/cycle.
This property is later used to dramatically reduce the size of the convolutional layers in hardware through the use of word and bit serial adders discussed in Section \ref{sec:conv}.

As a hardware block, the Max Pool is implemented in a pipelined way comparing the $k^2$ inputs over multiple cycles to determine the output for each component of the pixel.
When using the input configuration from Figure \ref{fig:imageWindow}, the output of the Max Pool is bursty, either producing results frequently on particular image rows or otherwise outputting nothing.
This requires the use of a \textit{FIFO} at the output of a Max Pool layer.

\subsection{Convolution}
\label{sec:conv}
Convolutional layers are where the bulk of the computation is performed in the network and hence require a large fraction of the total hardware.
Reducing the size of this hardware is essential to the viability of the method proposed in this paper.
As stated previously, this paper is based on the assumption that the convolution uses ternary weights known and fixed at design time.
The im2col transform described in Section \ref{sec:buffering} produces a dense vector to multiply into the convolutional weights matrix.
To understand how the convolutional layers are implemented in this paper, consider the example convolutional weights shown in Table \ref{table:convTriExample}.
This shows a single trinarised $3 \times 3$ convolutional filter on a grayscale image centered on an input pixel with value `e'.
The dense input vector of a $3 \times 3$ convolutional filter, represented by variables $a$-$i$ such as in Figure \ref{fig:imageWindow}, is multiplied by weights known at design time.
The output of this convolution operation is then given as $z_0=-a + c + e + f - h$.
A key insight of this paper is this equation can be implemented without multiplications as an adder tree in hardware allowing the removal of zero weights.

More generally, the convolution computes Equation \ref{equ:tnn_conv_unrolled} which is a sparse matrix vector operation where the matrix is ternary and known at design time.
The sparse matrix vector multiplication can be implemented as a row of multiplications followed by a different summation for each output.
The implementation of the convolution in this paper aims to exploit sparsity by removing any multiplications with zero, reducing the size or pruning the adder tree required to compute the summation.
The remaining values are multiplied by either $-1$ or 1.
Hence the multiplications can be removed entirely from the design with the 1 or $-1$ weight just indicating whether to add or subtract from the sum.
As there are typically numerous filters in the convolution, many summations are performed in parallel.
Using knowledge of the weights ahead of time it is possible to merge subexpressions within the summations to reduce hardware usage.

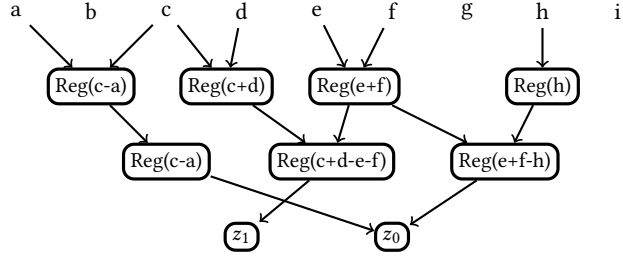
\begin{figure}
\centering
\begin{tikzpicture}

\tikzstyle{mybox} = [draw=black, very thick,
    rectangle, rounded corners, scale=0.9]

\node at (-6, 0 ) (a) {a};
\node at (-5, 0 ) (b) {b};
\node at (-4, 0 ) (c) {c};
\node at (-3, 0 ) (d) {d};
\node at (-2, 0 ) (e) {e};
\node at (-1, 0 ) (f) {f};
\node at (0, 0 ) (g) {g};
\node at (1, 0 ) (h) {h};
\node at (2, 0 ) (i) {i};

\node [mybox] at (-5, -1) (sub0) {Reg(c-a)};
\draw[->,thick] (a) -- (sub0);
\draw[->,thick] (c) -- (sub0);

\node [mybox] at (-1.5, -1) (add0) {Reg(e+f)};
\draw[->,thick] (e) -- (add0);
\draw[->,thick] (f) -- (add0);

\node [mybox] at (-3.2, -1) (add1) {Reg(c+d)};
\draw[->,thick] (c) -- (add1);
\draw[->,thick] (d) -- (add1);

\node [mybox] at (1, -1) (reg0) {Reg(h)};
\draw[->,thick] (h) -- (reg0);

\node [mybox] at (-4, -2) (reg1) {Reg(c-a)};
\draw[->,thick] (sub0) -- (reg1);

\node [mybox] at (0.5, -2) (sub1) {Reg(e+f-h)};
\draw[->,thick] (add0) -- (sub1);
\draw[->,thick] (reg0) -- (sub1);

\node [mybox] at (-1.8, -2) (sub2) {Reg(c+d-e-f)};
\draw[->,thick] (add1) -- (sub2);
\draw[->,thick] (add0) -- (sub2);

\node [mybox] at (-1, -3) (res) {$z_0$};
\draw[->,thick] (reg1) -- (res);
\draw[->,thick] (sub1) -- (res);

\node [mybox] at (-3, -3) (res2) {$z_1$};
\draw[->,thick] (sub2) -- (res2);

\end{tikzpicture}
\caption{Computing $z_0 = c + e + f - ( a + h )$ and $z_1 = c + d - e - f$}
\label{fig:convCompute}
\end{figure}

An example of how knowledge of the weights ahead of time can reduce the amount of hardware through subexpression elimination is shown in Figure \ref{fig:convCompute}. Here, an additional equation $z_1 = c + d - e - f$ is computed in parallel to $z_0$ simultaneously and sharing the subexpression $e + f$.

\begin{figure}
\centering
\begin{tikzpicture}

\tikzstyle{mybox} = [draw=black, very thick,
    rectangle, rounded corners]

\node (start) { Reset };
\node at (2.5, 2) (a) { a };
\node at (3.25, 2) (b) { b };

\node[draw=black, very thick,rectangle] at (3, 1) (sum) { + };
\node at (2, 1) (carry) { Carry };

\node[mybox] at (2, 0) (carryout) { Carry Out};
\node[mybox] at (4, 0) (out) {Out};

\draw[->,thick] (a) -- (sum);
\draw[->,thick] (b) -- (sum);
\draw[->,thick] (start) -- (carryout);
\draw[->,thick] (carry) -- (sum);
\draw[->,thick] (sum) -- (carryout);
\draw[->,thick] (carryout) -- (carry);
\draw[->,thick] (sum) -- (out);

\draw[->,thick] (out) -- (4,-1);

\end{tikzpicture}
\caption{Word or Bit Serial Adders/Subtractors}
\label{fig:serialAdder}
\end{figure}
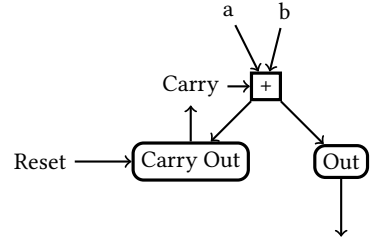

This approach requires a specialized datapath to compute the results.
The hardware implementation depends on $p$, the number of pixels arriving each cycle for a given layer.
Under the assumption $p\geq 1$ it is straight forward to construct an efficient hardware block by implementing $p$ versions of the adder trees in parallel.
However, when $p<1$ a single adder tree results in idle cycles and an inefficient design.
Time multiplexing the computation for different sets of weights destroys a lot of the advantages of this approach as it becomes much more difficult to eliminate subexpressions and exploit sparsity.
However, if the activations have a sufficently high bitwidth and based on $p$, word or bit serial adders can be used to implement the adder trees shown in Figure \ref{fig:serialAdder}.
Assuming 16 bit inputs, a parallel 16 bit adder requires 2 slices on a Xilinx Ultrascale FPGA to be implemented.
If this computation is performed over 4 cycles with a 4 bit word size, the area required to implement the adder is only $\frac{1}{2}$ of a slice.
With a bit serial adder, the result is calculated over 16 cycles and only requires a single LUT or $\frac{1}{8}$ of a slice.

\begin{table}
\centering
\caption{ An example $3 \times 3$ Convolutional Filter }
\label{table:convTriExample}
\begin{tabular}{c c c}
a $\times$~$(-1)$ & b $\times$~0      & c $\times$~1 \\
d $\times$~0      & e $\times$~1      & f $\times$~1 \\
g $\times$~0      & h $\times$~$(-1)$ & i $\times$~0 \\
\end{tabular}
\end{table}

The carry is reset to begin computation of the next set of numbers.
For addition, the carry is reset to 0 and for subtraction to 1.
The inputs $a$ and $b$ are added with the carry to output the result for that word or bit.
For subtraction, $b$ can be inverted inside the slice.
The carry out is stored for the next word or bit to be computed.
The output is passed directly into the next adder of the tree.
This reduces the hardware cost for implementing the layers while allowing them to match the required pixel rate $p$.

This approach can be applied generally to any 2D convolution with ternary weights and takes advantage of unstructured sparsity very effectively.
Sharing of results dramatically reduces the hardware required and is discussed further in Section \ref{sec:subexpression}.
The convolution operation outputs a vector based on the number of filters the layer contains which is typically followed by batch normalization (BN) and an activation functions such as ReLU.

\subsection{Scale and Shift}
\label{sec:scale_and_shift}
Batch Normalization (BN) is included in nearly every network due to its ability to improve training times and reduce overfitting. 
In feed-forward mode, it is reduced to a simple equation 
\begin{equation}
\mathbf{y} = \mathbf{a} \odot \mathbf{x} + \mathbf{b}
\label{equ:nn_batch_norm}
\end{equation}
where $\mathbf{x}$ represents activations, $\mathbf{a}$ and $\mathbf{b}$ are constants to scale and shift from the batch normalization operation known at design time and $\odot$ is the elementwise product.
Equation \ref{equ:nn_batch_norm} is referred to in this paper as a scale and shift operation.
As BN typically directly follows the convolutional layer, the scaling coefficient of the scaling $s$ from Equation \ref{equ:tnn_conv_unrolled} can be combined into a single equation
\begin{equation}
\mathbf{y} = \mathbf{c} \odot \mathbf{x} + \mathbf{b}
\label{equ:nn_scale_shift}
\end{equation}
where $\mathbf{c} = s\cdot \mathbf{a}$ as $\mathbf{a}$ and $s$ are constants known at design time.

The Scale and Shift hardware block can be implemented with a straight forward multiplication and addition using the constant values of $c,b$ for each output followed by an activation function.
This requires a fixed point multiplication with a constant after each layer, for each output filter.
This is achieved by having a multiplication operation at all outputs followed by an adder for the bias and finally having the activation function applied on $y$.
As the multiplications are with constants, depending on the value they might be implemented with or without DSPs at the discretion of the tools.

\subsection{MUX Layers}
\label{subsec:muxlyr}
Convolutional layers output a three dimensional image of size $W \times W \times D$ where a single pixel is a vector of size $D$ so each output only depends on a small section of the image.
Following multiple pooling layers, the convolution will output $D$ values every $M$ cycles.

The purpose of the MUX Layer hardware block is to transform a bursty output into a steady one.
For the convolution this block will take $D$ values every $M$ cycles as input and output $\frac{D}{M}$ values each cycle.
The MUX Layer implements this by buffering the vector of $D$ values in registers and making a multi-cycle MUX to output a portion of the values in order each cycle.
This reduces the width of the bus required to transport the values.

\subsection{Dense Layers}
\label{sec:dense}
CNNs can include dense layers at the end of the network for the final classification.
The output of the convolutional layers is flattened, then passed into the dense layer.
Unlike the convolutional layers the dense layer requires the entire image to be available before any of the outputs can be computed.
Similar to the rest of the network, it is assumed they also have ternary weights.

\begin{figure}
\centering
\begin{tikzpicture}

\tikzstyle{mybox} = [draw=black, very thick,
    rectangle, rounded corners, scale=0.9]

\node at (0, 0 ) (a) {a};
\node at (1, 0 ) (wa) {$w_a$};
\node at (2, 0 ) (b) {b};
\node at (3, 0 ) (wb) {$w_b$};
\node at (4, 0 ) (c) {c};
\node at (5, 0 ) (wc) {$w_c$};
\node at (6, 0 ) (d) {d};
\node at (7, 0 ) (wd) {$w_d$};

\node [mybox] at (0.5, -1) (mult0) {Reg(a*$w_a$)};
\draw[->,thick] (a) -- (mult0);
\draw[->,thick] (wa) -- (mult0);

\node [mybox] at (2.5, -1) (mult1) {Reg(b*$w_b$)};
\draw[->,thick] (b) -- (mult1);
\draw[->,thick] (wb) -- (mult1);

\node [mybox] at (4.5, -1) (mult2) {Reg(c*$w_c$)};
\draw[->,thick] (c) -- (mult2);
\draw[->,thick] (wc) -- (mult2);

\node [mybox] at (6.5, -1) (mult3) {Reg(d*$w_d$)};
\draw[->,thick] (d) -- (mult3);
\draw[->,thick] (wd) -- (mult3);

\node [mybox] at (1.5, -2) (add0) {Reg(a*$w_a$ + b*$w_b$)};
\draw[->,thick] (mult0) -- (add0);
\draw[->,thick] (mult1) -- (add0);

\node [mybox] at (5.5, -2) (add1) {Reg(c*$w_c$ + d*$w_d$)};
\draw[->,thick] (mult2) -- (add1);
\draw[->,thick] (mult3) -- (add1);

\node [mybox] at (3.5, -3) (add2) {Reg(a*$w_a$ + b*$w_b$ + c*$w_c$ + d*$w_d$)};
\draw[->,thick] (add0) -- (add2);
\draw[->,thick] (add1) -- (add2);

\node [mybox] at (3.5, -4) (add3) {Reg(a*$w_a$ + b*$w_b$ + c*$w_c$ + d*$w_d$ + sum)};
\draw[->,thick] (add2) -- (add3);
\draw[->,thick] (add3) -- (3.5, -5) -- (7.5, -5) -- (7.5, -4) -- (add3) node [midway, above] (TextNode) {sum};;

\end{tikzpicture}
\caption{Multiplying and Accumulating weights for the Dense Layer}
\label{fig:denseCompute}
\end{figure}
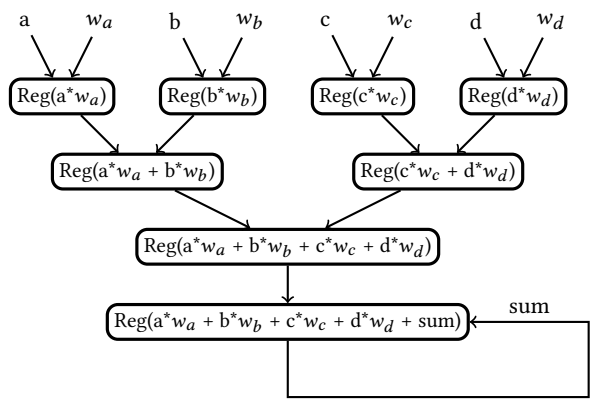

While the dense layer is also a matrix vector operation, it typically has a significantly larger number of weights meaning it is expensive to expand it into a tree as with the convolutional layers.
Additionally, as a dense layer is operated on the vector of the entire image, it would be inefficient to use the technique described for convolutional layers.
The main reason for this is the pruned adder tree approach would require the entire image to be buffered before beginning the computation.
This means the hardware to compute the dense layer only being utilised once each image and would be inefficient and unreasonably large in hardware.
For this reason, taking advantage of sparsity and common subexpression elimination (CSE) such as in Figure~\ref{fig:convCompute} is not efficient for dense layers.
Because of this, the approach common in the literature of streaming the weights from memory, multiplying them into the activations and accumulating the result is adopted as shown in Figure \ref{fig:denseCompute}.
As the weights are known ahead of time they can be stored in read only memory blocks on chip.
The number of MAC units as shown in Figure~\ref{fig:denseCompute}, depends on the size of the dense layer.
Each output of the dense layer has its own accumulation block.

\section{Subexpression Elimination}
\label{sec:subexpression}
This section discusses techniques that can be applied to merge common subexpressions in the implementation of the pruned adder trees to reduce the hardware required for implementation.
The subexpression elimination problem has a long history in computer science and its solvers are widely used in many applications such as the GCC compiler.
It has been determined to be NP-hard and hence can only be solved approximately \cite{cs84} for large problems.
The subexpression elimination discussed in this paper is a specialized case of the problem.
It is an unknown dense vector multiplied by a sparse matrix of constant values, restricted to $-1,0,1$.
Figure \ref{fig:convCompute} shows the desired solution to the problem given then input equations $z_0 = c + e + f - ( a + h )$ and $z_1 = c + d - e - f$.
The solution should be a pipelined adder tree aiming for the smallest implementation on an FPGA.
As an FPGA has registers after adder blocks, the cost for Add+Reg is considered the same as just Reg and should optimize this for minimal cost.
The implementation details of the Add+Reg block is considered irrelevant.
There are three related approaches to this problem in the literature.
RPAG by Kumm et al. \cite{kumm2017} uses constant integer values in the matrix.
Hsiao et al. \cite{hsiao2006memory} use common subexpression elimination (CSE) on a binary matrix for implementation of AES in hardware which will be modified for ternary values and referred to as top-down CSE (TD-CSE) in this paper.
Wu et al. \cite{wu2013improving} expand on this to what they call gate level delay computing CSE (GLDC-CSE) for binary values which will be modified for ternary values and referred to as bottom-up CSE (BU-CSE) in this paper.

The constant matrix optimization RPAG of Kumm et al. uses a graph-based algorithm \cite{kumm2012pipelined} to solve the generalized constant multiplication problem where the weights can be arbitrary integer numbers instead of restricted to $-1, 0, 1$.
Conversely, Hsiao et al. require values of either $0$ or $1$ to use TD-CSE and is hence a specialization of the problem required to be solved in this work.

\subsection{RPAG Algorithm}
Kumm et al. \cite{kumm2017} propose a method to perform subexpression elimination with integer values.
The algorithm looks at all the outputs of a matrix-vector multiplication and calculates the minimal tree depth, $d$, required to get the results.
At this depth, it then tries to determine the minimum number of terms needed at depth $d-1$ to compute the terms at depth $d$.
It then performs this iteratively until the depth is 1 resulting in the whole tree being generated.
These steps perform a broad search of the space and hence find very good solutions. 
However, common subexpressions are searched which can be shared between integer coefficients which is not necessary in our case. This and the broad search makes the approach computationally intensive and only suitable for relative small matrices.
The algorithm was recently extended to also utilize 3-input adders, which can be efficiently mapped to recent FPGAs \cite{Hardieck2018}.

\subsection{Top-Down CSE Approach}

The TD-CSE algorithm proposed by Hsiao et al. \cite{hsiao2006memory} iteratively eliminates subexpressions of size 2.
To explain the TD-CSE approach, consider the following example of a maxtrix-vector product:
\begin{align}
  \mathbf{y}=
  \left(
    \begin{array}{cccccc}
         0 & 0 & 1 & 1 & 0 & 0 \\
         1 & 0 & 1 & 1 & 1 & 0 \\
         0 & 1 & 0 & 0 & 1 & 1 \\
         0 & 1 & 0 & 0 & 0 & 1 \\
         1 & 0 & 1 & 1 & 0 & 0 \\
         1 & 0 & 0 & 1 & 0 & 0 \\
         0 & 1 & 0 & 0 & 1 & 1 \\
    \end{array}
  \right)
  \left(
    \begin{array}{cccccc}
         x_0\\
         x_1\\
         x_2\\
         x_3\\
         x_4\\
         x_5
    \end{array}
  \right)
  =
  \left(
    \begin{array}{cccccc}
         x_2 + x_3\\
         x_0 + x_2 + x_3 + x_4\\
         x_1 + x_4 + x_5\\
         x_1 + x_5\\
         x_0 + x_2 + x_3\\
         x_0 + x_3\\
         x_1 + x_4 + x_5\\
    \end{array}
  \right) \ .
  \label{eq:TD-CSE}
\end{align}
The TD-CSE algorithm counts the frequency of all subexpressions of size two, selecting the most frequent.
In this case, it is $x_2 + x_3$, occurring 3 times.
This expression is removed by defining $x_6 = x_2 + x_3$ and replacing all previous equations with $x_6$ resulting in 
\begin{align}
  \mathbf{y}=
  \left(
    \begin{array}{cccccc}
         x_6  \\
         x_0 + x_4 + x_6 \\
         x_1 + x_4 + x_5 \\
         x_1 + x_5 \\
         x_0 + x_6 \\
         x_0 + x_3 \\
         x_1 + x_4 + x_5 \\
    \end{array}
  \right) \ .
\end{align}
This process continues until there are no more subexpressions occurring more than once.




This approach can be thought of as building multiple adder trees from the inputs to the outputs by creating an adder each iteration.
There are some tricks to implement this approach that can dramatically improve the runtime.
When calculating the most common pairs the results can be stored and reused as, after removing a subexpression, most of these values do not need to be recalculated.
The first iteration requires the number of expressions containing each pair of variables to be computed where $n$ is the number of expressions, $v$ is the number of variables and $p$ is the number of pairs of variables in the expressions.
For example in Equation~\ref{eq:TD-CSE}, $n = 7$, $v=6$ and $p = {v \choose 2} = \frac{v(v-1)}{2}$.
The complexity of the first iteration is then $O(nv^2)$.
These results are then stored, and when a pair of variables are chosen to be removed as a common subexpression, the computation of the update is only $O(nv)$.
For the example in 
Equation~\ref{eq:TD-CSE}, after the subexpression is removed only combinations containing $x_2$, $x_3$ and $x_6$ would need to be computed.
The count of previous pairings between say, $x_0$ and $x_1$, do not need to be updated.
Another minor optimization to further skip computations is to look at combinations with $x_2$ and $x_3$.
If there were no expressions in common in the initial equations, say between $x_2$ and $x_5$, then the initial calculations would have a value of zero here.
As the number of subexpressions can only decrease after removing a subexpression, this value of zero does not need to be recomputed and the pair between $x_2$ and $x_5$ can be skipped.
As the problem grows in size, the amount this saves increases but the update is still $O(nv)$.
It should also be noted that $v$ grows by 1 each new subexpression that is removed as a new variable is created for it.
This method is generalized to $-1,0,1$ reasonably easily by creating another table for subexpressions with different signs.

\subsection{Bottom-Up CSE Approach}
Wu et al. \cite{wu2013improving} propose a method they term gate-level delay computing CSE (GLDC-CSE) for AES S-Box implementation.
We expand this method to $-1, 0, 1$ instead of $0,1$ and refer to it as bottom-up CSE (BU-CSE).
It considers building the tree from the other direction than TD-CSE as in RPAG.
Instead of starting at the inputs, it starts at the outputs and works back to the inputs.
Compared with TD-CSE, finding common expressions is more computationally intensive but can find better results as larger common subexpressions are preferred.
When building the tree from the bottom up, the size of the largest common subexpression needs to be determined for every pair of vectors.
The largest common subexpression is then selected to be removed.
For the example in Equation~\ref{eq:TD-CSE}, $x_0 + x_2 + x_3$ appears twice.
The subexpression, $x_6=x_0 + x_2 + x_3$ is added to the table of updated equations leading to
\begin{align}
  \mathbf{y}=
  \left(
    \begin{array}{cccccc}
         x_2 + x_3  \\
         x_4 + x_6 \\
         x_1 + x_4 + x_5 \\
         x_1 + x_5 \\
         x_6 \\
         x_0 + x_3 \\
         x_1 + x_4 + x_5 \\
         x_0 + x_2 + x_3 \\
    \end{array}
  \right) \ .
\end{align}
Note the new expression, $x_0 + x_2 + x_3$ added in the bottom row.
This is because $x_6$ still contains a subexpression of $x_2 + x_3$ or $x_0 + x_3$ that can be removed.


The algorithm for removing common terms is described as follows:
\begin{enumerate}
    \item Compute the number of common terms for each pair of vectors and store this as the \textit{pattern matrix}
    \item Find the largest value in the pattern matrix and the vectors it corresponds to
    \item Remove that subexpression from all matching vectors following the process described for the example in
    Equation~\ref{eq:TD-CSE}
    \item Update the \textit{pattern matrix}
    \item Go to step 2 until the largest value in the \textit{pattern matrix} is 1
\end{enumerate}

\section{Case Study on the VGG CNN of Li et al. for the CIFAR10 data set}
This section considers applying the above techniques to the network described in Li~et~al.~\cite{li2016ternary}, as applied to the CIFAR10 dataset which classifies 10 different image types: airplane, automobile, bird, cat, deer, dog, frog, horse, ship and truck.

\subsection{CNN Training}
A CNN similar to VGG-7 was trained using a scheme adopted from Li et al. 
The following parameters were used:
\begin{itemize}
\item A batch size of 128
\item An initial learning rate of 0.1
\item Learning rate decays by a factor of 0.1 every 100 epochs.
\item Run for 120k steps or 307.2 epochs
\end{itemize}
We also used data augmentation as described by Li et al. \cite{li2016ternary} where the image is padded with 4 pixels to each side and randomly cropped back to the 32x32 image size during training.
Li et al.'s method of training a CNN utilises a threshold
\begin{equation}
\Delta^* \approx \epsilon \cdot E(|W|)
\label{equ:li_et_al_thres}
\end{equation}
where Li recommend a value of $\epsilon = 0.7$.
If the magnitude of the floating point weight is less than this threshold, $\Delta^*$, it is set to 0, otherwise the sign of the floating point weight is used multiplied with a scaling factor $s$.
The parameter $\epsilon$ thus directly controls the sparsity of a layer in this method.

This technique of adjusting $\epsilon$ to tradeoff accuracy and sparsity of the network simply and quickly is not explored in Li et al. \cite{li2016ternary} but is a strong feature of their method and important to our approach as sparsity directly effects the size of the hardware.
The network architecture proposed by Li et al. was modified by reducing the number of filters by half in all convolutional layers of the network and the size of the dense layer by $8\times$ before training.
This was done as it had a minor impact on accuracy but dramatically reduces the size of the network for a hardware implementation.

The results obtained for different values of $\epsilon$ are summarised in Table~\ref{table:cnnPerf}.
We also note that the result of our implementation in the second row of the table is higher than the 92.6\% reported by Li et al. \cite{li2016ternary} which is likely due to longer training time.
As Table~\ref{table:cnnPerf} shows, the choice of $\epsilon = 0.7$ produces a network with high accuracy but is relatively dense, with a sparsity of around 47\%.
The remaining results were obtained by maintaining $\epsilon = 0.7$ in the first layer, $\epsilon = 1.0$ for the dense layers and adjusting all the other convolutional layers to use a different value of $\epsilon$.
As Equation \ref{equ:li_et_al_thres} shows, a larger value of $\epsilon$ corresponds to a higher threshold.
As the weight is set to 0 if it is below this threshold, a higher value of $\epsilon$ means increased sparsity.
As $\epsilon$ increases, a slight drop in accuracy is observed with an increase in sparsity dependent on the value shown in Table~\ref{table:cnnPerf}.

With the value of $\epsilon = 1.4$ chosen from Table~\ref{table:cnnPerf}, the computation required for the network is reduced by almost half compared to a value of $\epsilon = 0.7$.
This is extremely advantageous as, for our implementation of Equation \ref{equ:tnn_conv_unrolled}, increased sparsity reduces the area allowing a larger design to fit on a FPGA.
The final network architecture chosen is given in Table~\ref{table:cnnArch}.
Batch normalization and ReLU activations are used after each convolutional and dense layer.

\begin{table}
\centering
\caption{Effect of $\epsilon$ on sparsity and accuracy for CIFAR10}
\label{table:cnnPerf}
\begin{tabular}{l l l l}
\toprule
TNN Type & $\epsilon$ & Sparsity ( \% ) & Accuracy \\
\midrule
Graham \cite{graham2014fractional} (Floating Point) & - &  - & 96.53\% \\
Li et al. \cite{li2016ternary}, full-size & 0.7 & $\approx 48$ & 93.1\% \\
Half-size & 0.7 & $\approx 47$ & 91.4\% \\
Half-size & 0.8 & $\approx 52$ & 91.9\% \\
Half-size & 1.0 & $\approx 61$ & 91.7\% \\
Half-size & 1.2 & $\approx 69$ & 91.9\% \\
Half-size & 1.4 & $\approx 76$ & 90.9\% \\
Half-size & 1.6 & $\approx 82$ & 90.3\% \\
Half-size & 1.8 & $\approx 87$ & 90.6\% \\
\bottomrule
\end{tabular}
\end{table}

\begin{table}
\centering
\caption{Architecture of the CNN used in this paper}
\label{table:cnnArch}
\begin{tabular}{l l l l l}
\toprule
Layer Type & Input Image Size & Num Filters & $\epsilon$ & Sparsity\\
\midrule
Conv2D & $32 \times 32 \times 3$ & 64 & 0.7 & 54.7\% \\
Conv2D & $32 \times 32 \times 64$ & 64 & 1.4 & 76.9\% \\
Max Pool & $32 \times 32 \times 64$ & 64 & - & -\\
Conv2D & $16 \times 16 \times 64$ & 128 & 1.4 & 76.1\% \\
Conv2D & $16 \times 16 \times 128$ & 128 & 1.4 & 75.3\% \\
Max Pool & $16 \times 16 \times 128$ & 128  & - & - \\
Conv2D & $8 \times 8 \times 128$ & 256 & 1.4 & 75.8\% \\
Conv2D & $8 \times 8 \times 256$ & 256 & 1.4 & 75.4\% \\
Max Pool & $8 \times 8 \times 256$ & 256 & - & - \\
Dense & 4096 & 128 & 1.0 & 76.2\% \\
Softmax & 128 & 10  & 1.0 & 58.4\% \\
\bottomrule
\end{tabular}
\end{table}

\subsection{The Impact of Subexpression Elimination Techniques}
To compare the effectiveness of the three CSE techniques, ten models with the above network, each with a random initialisation, were trained with a chosen $\epsilon = 1.4$.
The accuracy obtained from these models with floating point activations is $91.7\% \pm 0.1\%$ and using 16 bit fixed point activations is $90.9\% \pm 0.1\%$.
The first layer was selected from all ten trained models with a matrix size of 27 inputs and 64 outputs where all values are $-1,0,1$.
The average results of the techniques are compared in Table~\ref{tab:comparison_cse} for 2-input as well as 3-input adders.

\begin{table}
    \centering
    \caption{Comparison of CSE techniques on ternary weights with 2-input adders (top) and 3-input adders (bottom) for the first layer}
    \begin{tabular}{lcccc}
         \toprule
         Technique & Avg Adders & Avg Reg & Avg Add/Reg & Avg Time (s)\\
         \midrule
         None (2-input)   &  755.2 & 146.5 & 901.7 & - \\
         RPAG (2-input)   & 487.3 & 26.7 & \textbf{ 460.6 } &  52.3 \\
         TD-CSE (2-input) & 300.4 & 309.8 & 610.2 & 0.317 \\
         BU-CSE (2-input) & 296.3 & 345.7 & 642 & 0.459 \\
         \cmidrule(rl){1-5}
         None (3-input)   & 406.2 & 54.1 & 460.3 & - \\
         RPAG (3-input)   & 332 & 7.5 & \textbf{324.5} & 28460.8 \\
         TD-CSE (3-input) & 246.7 & 294.5 & 541.2 & 0.297 \\
         BU-CSE (3-input) & 258.8 & 321.7 & 580.5 & 0.474 \\
         \bottomrule
    \end{tabular}
    \label{tab:comparison_cse}
\end{table}

It can be seen that RPAG has the lowest register and adder combined cost.
This is likely the smallest in hardware on an FPGA due to the structure of adders having optional registers at the outputs.
The cost on the FPGA of an adder and a register compared with just a register is similar, hence the rightmost column provides the most meaningful comparison \cite{kumm2012pipelined}. 
These results suggest that RPAG is the best technique, followed by TD-CSE, then BU-CSE.

However, experiments showed that RPAG does not scale well to the larger layers as shown by the significant running time for the smallest layer in Table \ref{tab:comparison_cse}.
The scalability of these techniques to larger matrices is significant for this application, as the matrix size grows significantly in the later layers.
For those, the TD-CSE method is the most scalable as it can skip large portions of the computation each cycle.
These results only reflect a first layer of a network with more outputs than inputs.
In all subsequent layers there are more inputs than outputs in the rest of a CNN.

\begin{table}
    \centering
    \caption{Comparison of CSE techniques on ternary weights with 2-input adders (top) and 3-input adders (bottom) for the second layer }
        \begin{tabular}{lcccc}
        \toprule
         Technique & Avg Adders & Avg Reg & Avg Add/Reg & Avg Time (s) \\
         \midrule
         None (2-input) & 8936.4 & 254.3 & 9190.7 & - \\
         TD-CSE (2-input) & 3970.6 & 1521.5 & 5492.1 & \textbf{20.68} \\
         BU-CSE (2-input) & 3890.3 & 902.0 & 4792.3 & 55.94 \\
         \cmidrule(rl){1-5}
         None (3-input) & 4536.7  & 91.6 & 4628.3 & - \\
         TD-CSE (3-input) & 2826.8 & 1447.9 & 4274.7 & 20.78 \\
         BU-CSE (3-input) & 2425.6 & 439 & \textbf{2864.6} & 58.59 \\
         \bottomrule
    \end{tabular}
    \label{tab:comparison_cse_large_sparse}
\end{table}

Table \ref{tab:comparison_cse_large_sparse} shows a comparison between the TD-CSE and BU-CSE techniques for the larger matrices in layer 2.
These matrices had 576 inputs and 64 outputs with a sparsity of around $75$\%.
These results show that the BU-CSE technique finds better solutions for this configuration.
It was not possible to run RPAG on these matrices as the large size makes it infeasible.
This suggests that BU-CSE is better for larger matrices but the results are similar enough that both should be run to find a better solution.

\begin{figure}
\centering
\begin{tikzpicture}

\tikzstyle{mybox} = [draw=black, very thick,
    rectangle, rounded corners, scale=0.9]

\node[mybox] at (0, 2.5) (input) {Input Image};
\node[mybox] at (0, 1.25) (img) {Conv1/2 (16-bit adders)};
\node[mybox] at (0, 0) (mp1) { Max Pool 1 };
\node[mybox] at (0, -1.25) (conv3) {Conv3/4 (4-bit adders)};
\node[mybox] at (0, -2.5) (mp2) { Max Pool 2 };
\node[mybox] at (0, -3.75) (conv5) {Conv5/6 (1-bit adders)};
\node[mybox] at (0, -5) (mp3) { Max Pool 3 };
\node[mybox] at (0, -6.25) (mux) { MUX Layer };
\node[mybox] at (0, -7.5) (out) { Output };

\draw[->,thick] (input) --  (img) node [midway, left] (TextNode) {32x32x3} node [midway, right] (TextNode) {3 values every cycle};
\draw[->,thick] (img) --  (mp1) node [midway, left] (TextNode) {32x32x64} node [midway, right] (TextNode) {64 values every cycle};
\draw[->,thick] (mp1) --  (conv3) node [midway, left] (TextNode) {16x16x64} node [midway, right] (TextNode) {64 values every 4 cycles};
\draw[->,thick] (conv3) --  (mp2) node [midway, left] (TextNode) {16x16x128} node [midway, right] (TextNode) {128 values every 4 cycles};
\draw[->,thick] (mp2) --  (conv5) node [midway, left] (TextNode) {8x8x128} node [midway, right] (TextNode) {128 values every 16 cycles};
\draw[->,thick] (conv5) --  (mp3) node [midway, left] (TextNode) {8x8x256} node [midway, right] (TextNode) {256 values every 16 cycles};
\draw[->,thick] (mp3) --  (mux) node [midway, left] (TextNode) {4x4x256} node [midway, right] (TextNode) {256 values every 64 cycles};
\draw[->,thick] (mux) --  (out) node [midway, left] (TextNode) {4096} node [midway, right] (TextNode) {4 values every cycle};

\end{tikzpicture}
\caption{Impact of Max Pool Layers}
\label{fig:mpThroughput}
\end{figure}
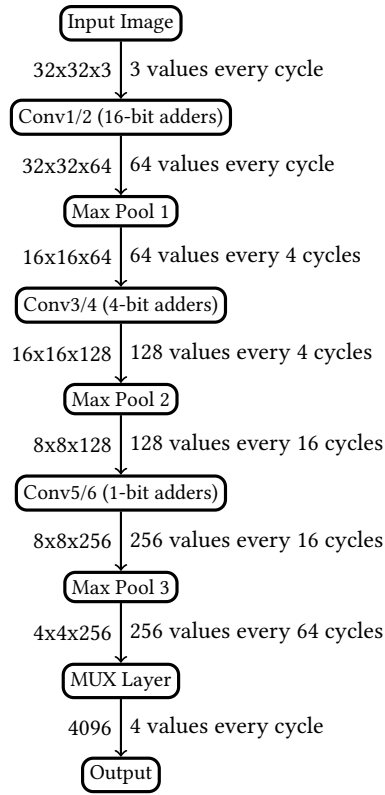

In the following, we selected one out of the ten models trained for $\epsilon=1.4$ and applied CSE to its weights in all convolutional layers to reduce the amount of computation required for the FPGA implementation.
Table \ref{tab:conv_cse} shows the results of running subexpression elimination on the convolutional layers of the trained network described in Table~\ref{table:cnnArch}, comparing different techniques.
The table does not show RPAG for layers 2-6 as it becomes computationally prohibitive.
The reason that only the first two layers use 3-input adders in Table~\ref{table:cnnArch} is that the remaining layers compute the result over multiple cycles as discussed with Figure~\ref{fig:mpThroughput}. Here, 3-input adders can not be efficiently utilized.

\begin{table}
    \centering
    \caption{Common Subexpression Elimination and Hardware Usage on the Convolutional Layers \\
    The adder-tree section of the convolution was run out-of-context with area optimizations; Only 2-input adds are compared in this table for consistency}
    \label{tab:conv_cse}
    \begin{tabular}{lcccccccccc}
        \toprule
         Layer & Method & Adds & Regs & Adds+Regs & Time(s) & Mem(GB) & CLB/148K & FF/2.4M & LUTS/1.2M & P\&R(hrs) \\
         \midrule
         \multirow{4}{*}{1}
         & \multicolumn{1}{c}{ None } & \multicolumn{1}{c}{ 731 } & \multicolumn{1}{c}{ 137 }& \multicolumn{1}{c}{ 868 } & - & - & 1400 & 8723 & 8272 & 0.5  \\
         & \multicolumn{1}{c}{ RPAG } & \multicolumn{1}{c}{ 451 } & \multicolumn{1}{c}{ 31 }& \multicolumn{1}{c}{ \textbf{ 482 } } & 64 & 0.008 & 894 & 5764 & 6260 & 0.48 \\
         & \multicolumn{1}{c}{ TD-CSE } & \multicolumn{1}{c}{ 295  } & \multicolumn{1}{c}{ 304 }& \multicolumn{1}{c}{ 599 } & 0.4 & 0.029 & - & - & - & - \\
         & \multicolumn{1}{c}{ BU-CSE } & \multicolumn{1}{c}{ 295 } & \multicolumn{1}{c}{ 321 }& \multicolumn{1}{c}{ 616 } & 0.5 & 0.03 & 820 & 4499 & 5230 & 0.45 \\
         \cmidrule(rl){1-11}
         \multirow{3}{*}{2}
         & \multicolumn{1}{c}{ None } & \multicolumn{1}{c}{ 8432 } & \multicolumn{1}{c}{ 249 }& \multicolumn{1}{c}{ 8681 } & - & - & 15231 & 119848 & 116345 & 1.08\\
         & \multicolumn{1}{c}{ TD-CSE } & \multicolumn{1}{c}{ 3782 } & \multicolumn{1}{c}{ 1517 }& \multicolumn{1}{c}{ 5299 } & 24 & 0.1 & - & - & - & -\\
         & \multicolumn{1}{c}{ BU-CSE } & \multicolumn{1}{c}{ 3686 } & \multicolumn{1}{c}{ 858 }& \multicolumn{1}{c}{ \textbf{4544} }& 64 & 0.17 & 10258 & 71908 & 66131 & 0.93 \\
         \cmidrule(rl){1-11}
         \multirow{3}{*}{3}
         & \multicolumn{1}{c}{ None } & \multicolumn{1}{c}{ 17481 } & \multicolumn{1}{c}{ 491 }& \multicolumn{1}{c}{ 17972 } & - & - & 15171 & 102657 & 77743 & 1.9 \\
         & \multicolumn{1}{c}{ TD-CSE } & \multicolumn{1}{c}{ 8466 } & \multicolumn{1}{c}{ 2299 }& \multicolumn{1}{c}{ 10765 } & 89 & 0.18 & - & - & - & - \\
         & \multicolumn{1}{c}{ BU-CSE } & \multicolumn{1}{c}{ 8492 } & \multicolumn{1}{c}{ 1878 }& \multicolumn{1}{c}{ \textbf{10370} } & 545 & 1.1 & 8772 & 61965 & 36611 & 1.13 \\
         \cmidrule(rl){1-11}
         \multirow{3}{*}{4}
         & \multicolumn{1}{c}{ None } & \multicolumn{1}{c}{ 36155 } & \multicolumn{1}{c}{ 586 }& \multicolumn{1}{c}{ 36741 } & - & - & 30536 & 206940 & 164458 & 4.25 \\
         & \multicolumn{1}{c}{ TD-CSE } & \multicolumn{1}{c}{ 17143 } & \multicolumn{1}{c}{ 4214 }& \multicolumn{1}{c}{ 21357 } & 873 & 0.63 & - & - & - & - \\
         & \multicolumn{1}{c}{ BU-CSE } & \multicolumn{1}{c}{ 17309 } & \multicolumn{1}{c}{ 3056 }& \multicolumn{1}{c}{ \textbf{20365} } & 2937 & 6.6 & 16909 & 118476 & 73581 & 2.68 \\
         \cmidrule(rl){1-11}
         \multirow{3}{*}{5}
         & \multicolumn{1}{c}{ None } & \multicolumn{1}{c}{ 71050 } & \multicolumn{1}{c}{ 1198 }& \multicolumn{1}{c}{ 72248 } & - & - & 18414 & 165794 & 85743 & 3.86 \\
         & \multicolumn{1}{c}{ TD-CSE } & \multicolumn{1}{c}{ 32829 } & \multicolumn{1}{c}{ 6830 }& \multicolumn{1}{c}{ 39659 } & 3088 & 1.2 & - & - & - & - \\
         & \multicolumn{1}{c}{ BU-CSE } & \multicolumn{1}{c}{ 33026 } & \multicolumn{1}{c}{ 6109 }& \multicolumn{1}{c}{ \textbf{39135} } & 25634 & 44 & 7579 & 89820 & 39805 & 1.72 \\
         \cmidrule(rl){1-11}
         \multirow{3}{*}{6}
         & \multicolumn{1}{c}{ None } & \multicolumn{1}{c}{ 144813 } & \multicolumn{1}{c}{ 1270 }& \multicolumn{1}{c}{ 146083 } & - & - & 35117 & 335134 & 180402 & 11.15 \\
         & \multicolumn{1}{c}{ TD-CSE } & \multicolumn{1}{c}{ 62653 } & \multicolumn{1}{c}{ 13852 }& \multicolumn{1}{c}{ 76505 } & 26720 & 4.8 & - & - & - & - \\
         & \multicolumn{1}{c}{ BU-CSE } & \multicolumn{1}{c}{ 63832 } & \multicolumn{1}{c}{ 10103 }& \multicolumn{1}{c}{ \textbf{73935} } & 147390 & 191.0 & 13764 & 160634 & 74696 & 3.08 \\
         \bottomrule
    \end{tabular}
\end{table}

\begin{table}
    \centering
    \caption{Improvement in resource usage when applying BU-CSE vs None}
    \label{tab:vivado_cse}
    \begin{tabular}{lcccc}
        \toprule
         Layer & \% decrease in Adds+Regs & \% decrease in CLBs & \%decrease in FFs & \% decrease in LUTs \\
         \midrule
         1 & -29.0 & -41.4 & -48.4 & -36.8 \\
         2 & -47.7 & -32.6 & -40.0 & -43.2 \\
         3 & -42.3 & -42.1 & -39.6 & -52.9 \\
         4 & -44.6 & -44.6 & -42.3 & -55.3 \\
         5 & -45.8 & -58.8 & -45.8 & -53.6 \\
         6 & -49.4 & -60.8 & -52.1 & -58.6 \\
        \bottomrule
    \end{tabular}
\end{table}

The results from Table \ref{tab:comparison_cse} show that RPAG is the most successful on the first layer of the network.
Table \ref{tab:conv_cse} shows that, except for the first layer, BU-CSE finds the superior solution on the chosen network.
One possible reason is the relative numbers of inputs and outputs as shown by the contrasting results in Tables ~\ref{tab:comparison_cse} and ~\ref{tab:comparison_cse_large_sparse}.
Layer~1 matrices used for the comparison in Table~\ref{tab:comparison_cse} have a smaller number of inputs than outputs.
Conversely, layers 2-6 have a much larger number of inputs than outputs.
The scalability of BU-CSE and TD-CSE should be mentioned as the time taken to run BU-CSE on the 6th (largest) convolutional layer is over a day, where as TD-CSE only needs a couple of hours and significantly less memory.
This time is relatively insignificant for the model updated reasonably infrequently and can be considered an extension of the training time.

The effectiveness of the CSE results can be compared by running Vivado with aggressive area optimizations in synthesis and implementation in out of context mode.
The Verilog generated without CSE (`None' in Table~\ref{tab:conv_cse}) is compared to the Verilog generated after optimization with the BU-CSE algorithm.
The effectiveness of CSE is compared with the difference in the originally generated Verilog in columns `Adds', `Regs' and `Adds+Regs'.
The runtime and peak memory used to run the CSE algorithm is shown in columns `Time' and `Mem'.
The last four columns show the logic resources after Place and Route  (P\&R) as well as its runtime. Table~\ref{tab:vivado_cse} compares the rows for None and BU-CSE from Table~\ref{tab:conv_cse} by calculating the decrease when using CSE. The results demonstrate that the solution to the CSE problem of an entire network layer is difficult for Vivado to manage.
The reason for the difference in results may be the abstracted view of the problem that the CSE methods have.
This allows them to explore the problem in more detail and find a better solution than Vivado.
Due to the use of word and bit serial adders, the amount of area that is used for layers 3-6 is reduced significantly compared to the amount of adders required.
As discussed previously in Section~\ref{sec:dense}, it is ineffective to use the pruned adder tree for the dense layers and thus CSE is not used for these.

\begin{table}
\centering
\caption{FPGA CNN Architecture blocks }
\label{table:fpgaBlocks}
\begin{tabular}{l c c c}
\toprule
Operation & Image Size In & Channel In & Channel Out \\
\midrule
Buffer & 32x32 & 3 & 3 \\
Conv & 32x32 & 3 & 64 \\
Scale and Shift & 32x32 & 64 & 64 \\
Buffer & 32x32 & 64 & 64 \\
Conv & 32x32 & 64 & 64 \\
Scale and Shift & 32x32 & 64 & 64 \\
Buffer & 32x32 & 64 & 64 \\
Max Pool & 32x32 & 64 & 64 \\
Buffer & 16x16 & 64 & 64 \\
Conv & 16x16 & 64 & 128 \\
Scale and Shift & 16x16 & 128 & 128 \\
\textbf{Buffer} & 16x16 & 128 & 128 \\
\textbf{Conv} & 16x16 & 128 & 128 \\
Scale and Shift & 16x16 & 128 & 128 \\
Buffer & 16x16 & 128 & 128 \\
Max Pool & 16x16 & 128 & 128 \\
Buffer & 8x8 & 128 & 128 \\
Conv & 8x8 & 128 & 256 \\
Scale and Shift & 8x8 & 256 & 256 \\
Buffer & 8x8 & 256 & 256 \\
Conv & 8x8 & 256 & 256 \\
Scale and Shift & 8x8 & 256 & 256 \\
Buffer & 8x8 & 256 & 256 \\
Max Pool & 8x8 & 256 & 256 \\
FIFO & 4x4 & 256 & 256 \\
MuxLayer & 4x4 & 256 & 4096 \\
Dense & 1x1 & 4096 & 128 \\
Scale and Shift & 1x1 & 128 & 128 \\
MuxLayer & 1x1 & 128 & 128 \\
Dense & 1x1 & 128 & 10 \\
\bottomrule
\end{tabular}
\end{table}

\subsection{FPGA Implementation}
The network described by Li et al. \cite{li2016ternary} contains floating point activations which limits FPGA performance as it is hardware intensive to implement add and multiply blocks.
After the network was trained, the batch normalization variables and scaling coefficients for the convolutional layers were extracted from the model.
A Python script then computed the network performance using these weights on the CIFAR10 test set in fixed point.
By outputting the maximum absolute values obtained at various stages of the calculation, it was determined to use a total of 16 bits to maintain accuracy, ensure no overflow and to simplify the implementation of the word and bit serial adders discussed in Section \ref{sec:conv}.
The weights of the network are all ternary values, however for the activations of the network, 4 fractional bits were used leaving 12 integer bits.
Each layer is followed by batch normalization and also has a scaling factor which are floating point values.
These constant values were combined in floating point, as discussed in Section~\ref{sec:scale_and_shift} and Equation \ref{equ:nn_scale_shift}, then quantised with 6 fractional bits leaving 10 integer bits.
The number of fractional bits was obtained experimentally to achieve maximum precision without the risk of overflow.
This was relatively simple to apply to the network and its impact on the performance 
achieves the same result as the floating point scaling reported in Table~\ref{table:cnnPerf}.
The design was then implemented using Chisel3 \cite{bachrach2012chisel} which facilitated the generation of irregular adder trees.
The CNN was partitioned into hardware blocks discussed in previous sections to stream the images through the network as shown in Table~\ref{table:fpgaBlocks}.

As discussed in Section~\ref{sec:conv}, convolutional layers can be implemented with word or bit serial adders if the throughput is low enough.
This is possible to apply to the 3rd and 4th convolutional layers as the Max Pool layer shrinks the image from $32\times 32$ to $16 \times 16$, dropping the throughput to $\frac{1}{4}$ of what it was originally.
As 16 bit fixed point is used for the activations the adder trees for the 3rd and 4th convolutional layers can use 4 bit word serial adders requiring $\frac{1}{4}$ of the area.
Similarly for the 5th and 6th convolutional layers, the throughput required is only one output every $\frac{1}{16}$ cycles.
Hence, bit serial adders are used to compact the design to $\frac{1}{16}$ of the area of a full adder while maintaining sufficient throughput to ensure a fully pipelined design.
This avoids the adders idling by creating hardware to compute the result over numerous cycles.

Figure \ref{fig:mpThroughput} demonstrates how this impacts the design at a high level.
The left side shows the image size, while the right side shows the throughput required.
This structure allows the pruned adder tree to be implemented efficiently while still exploiting unstructured sparsity and common subexpressions in the design to reduce the area required.
It is ineffective to implement the dense layer using CSE as discussed in Section \ref{sec:dense}.

\subsection{The Aggregated Network}
The VGG-7 network blocks for the FPGA design are described in Table~\ref{table:fpgaBlocks}.
This architecture is designed to accept a single pixel each cycle such that $p = 1$.
Given an image size of $W \times W$, $W^2$ cycles are therefore required to stream each image in.
Hence, classifying a new image every $W^2$ cycles and assuming a clock frequency of $f_\text{clk}$ gives a throughput of $\frac{f_\text{clk}}{W^2}$ classifications/second.
The ternary MACs required to compute the VGG-7 style network described in Table~\ref{table:cnnArch} for a single image is given in Table~\ref{table:ops}.
Operations in Max Pool and batch normalization are considered trivial and not counted in this total.

\begin{table}
\caption{Ops needed to compute}
\begin{tabular}{lllll}
\toprule
Layer & Num Mults & Num Mults & With Sparsity & With CSE \\
\midrule
Conv1 & 32*32*3*3*3*64 & 1769472 & 716800 & 630784 \\
Conv2 & 32*32*3*3*64*64 & 37748736 & 8637440 & 4653056 \\
Conv3 & 16*16*3*3*64*128 & 18874368 & 4559616 & 2654720 \\
Conv4 & 16*16*3*3*128*128 & 37748736 & 9396480 & 5213440 \\
Conv5 & 8*8*3*3*128*256 & 18874368 & 4656768 & 2504640 \\
Conv6 & 8*8*3*3*256*256 & 37748736 & 9356736 & 4731840 \\
Dense & 4096*128 & 524228 & 524228 & 1048456$^1$ \\
SM & 128*10 & 1280 & 1280 & 2560$^1$ \\
Total & 153289924 & 153\text{ MMACs/Image } & 38\text{ MMACs/Image } & 21 \text{ MOps/Image } \\
\bottomrule
\end{tabular}
\caption*{$^1$ Obtained by converting one MACs to two Ops}
\label{table:ops}
\end{table}

Hence for a fully floating point implementation with the equivalent throughput to this FPGA implementation, this network requires a total of $153\text{\,MMAC/Image}\times\frac{f_\text{clk}}{W^2}\text{ Images/sec}\times2\text{ Ops/MAC}$.
In the case of a network with ternary weights, the multiplication is with -1,0 and 1.
Table~\ref{table:ops} then shows the MMACs after accounting for sparsity, reducing the cost per image to 38\,MMACs.
As discussed in Section~\ref{sec:conv}, this can be implemented with only adds and subtracts with the CSE structure.
The multiplication is no longer performed in the convolutional layers but still in the dense layers and hence the final column in Table~\ref{table:ops} shows the MOps needed on the FPGA for a single image, counting 1 MAC as 2 Ops.
The actual computation performed on the FPGA with CSE is therefore $21\text{\,MOps/Image}\times\frac{f_\text{clk}}{W^2}\text{ Images/sec}$.
With these parameters and the implementation of the dense layers described in section \ref{sec:dense}, the largest dense layer needs $4096 \times 128 \times 2\text{\,b} = 1\text{\,Mb}$ of storage for the weights as each weight is only 2 bits.
As this is computed at a rate of 4 per cycle, the bandwidth for this memory is required to be $4 \times 128 \times 2\text{\,b} = 1\text{\,Kb}$ each cycle.
Given that a BRAM on a Xilinx device typically has an output bandwidth of 64 bits each cycle, this implies just 16 BRAMs are needed to store the weights, each storing 64\,kb of data.

\section{Results}
\label{se:results}
The entire system is implemented on an AWS F1 instance and achieves a frequency of 125\,MHz.
The entire system is a loopback application which passes the CIFAR10 images in from the C application using DPDK libraries \cite{dpdk}, onto the FPGA via PCIe, through the network and back to the C program.
The relatively low clock frequency of 125\,MHz was necessary due to routing congestion that we observed when using tighter clock constraints.
The bottleneck is routing between the convolutional layers.
Due to the windowing for the convolution in the buffer layer, a large amount of routing to different CLBs for the input of the matrix multiplication is required.
Due to the limits in routing resources, it is difficult to fanout the wide bus to all adders that require that input.
The critical path is in the layers that have high numbers of inputs that are going to a variety of locations.
Conv2 has 9*64 inputs with 16 bit bus width.
This is a wide bus width, but requires a smaller number of unique locations and hence the granularity of the routing can be course.
Conv4 has 9*128 inputs with 4 bit bus width and Conv6 has 9*256 inputs with 1 bit bus width.
With a larger convolutional layer, more fanout is required at these points.
The critical path in this design is between the two modules in bold in Table \ref{table:fpgaBlocks}.

\begin{table}
\caption{Vivado size hierarchy}
\begin{tabular}{lll}
\toprule
Block & LUTs/1182240 & FFs/2364480 \\ 
\midrule
Conv1 & 3764 ( 0.3\% )  & 10047 ( 0.4\% )\\ 
Conv2 & 40608 ( 3.4\% ) & 71827 ( 3.0\% )\\ 
Conv3 & 55341 ( 4.7\% )& 56040 ( 2.4\% )\\ 
Conv4 & 111675 ( 9.4\% )& 110021 ( 4.7\% ) \\ 
Conv5 & 73337 ( 6.2\% )& 79233 ( 3.4\% )\\ 
Conv6 & 127932 ( 10.8\% )& 139433 ( 5.9\% )\\ 
All Conv & 535023 ( 45.3\% )& 631672 ( 26.7\% ) \\ 
Dense & 12433 ( 1.1\% )& 19295 ( 0.8\% )\\ 
SM & 500 ( 0.04\% )& 442 ( 0.02\% )\\ 
Whole CNN & 549358 ( 46.5\% )& 659252 ( 27.9\% ) \\ 
Whole design & 787545 ( 66.6\% ) & 984443 ( 41.6\% ) \\ 
\bottomrule
\end{tabular}
\label{table:vivado_size}
\end{table}

The synthesisable Verilog register transfer level (RTL) code was generated in Chisel3, with custom modules being used to invoke components from the Vivado IP Core, such as the FIFOs.
The design was developed using Vivado 2018.2.
The hardware was verified with an input image passed into the design and obtaining identical output as verified against a Python script used to compute the fixed point performance of the network in Table~\ref{table:cnnPerf}.
The code for the implementation is publicly available on github\footnote{github.com/da-steve101/aws-fpga.git and github.com/da-steve101/binary\_connect\_cifar.git}.

Table \ref{table:vivado_size} shows the resources used by various layers of the network in the implemented design. In total, 66.6\% and 41.6\% of the LUT and FF resources were utilized.

For $f_\text{clk} = 125$\,MHz, the total theoretical Ops required for an equivalent dense floating point implementation is $153\text{\,MMAC/Image}\times\frac{125 MHz }{32^2}\text{ Images/sec}\times2\text{ Ops/MAC}$ = 37.3\,TOps/sec.
For this implementation, multiplications in the convolutional layers are not required and a significant portion of the operations can be removed.
Hence, in practice only approximately $2.5 \times 10^{12}$ Adds/sec are necessary.
The implemented system achieves the throughput of classifying 122k images/sec and a latency of 29\,\textmu s as only a fraction of the PCIe bandwidth is needed.
This is including a DPDK \cite{dpdk} virtual ethernet interface to send and receive packets from the FPGA.
Table~\ref{table:hardwareComparison} gives a comparison of this work with previously reported results for CIFAR10.
It shows the properties of different implementations.
There is a TOps column in this table to show the actual Ops computed on the FPGA, the logical Ops and the equivalent floating point Ops for the same sized network.
As these operations are typically very few bits, multiplying and accumulating with 1 or 2 bit numbers are typically integrated in a single logical statement.
For this reason the actual logical operations are also shown.
The values in order for the TOps column are Actual/Logical/Equivalent (A/L/E).
Table~\ref{table:hardwareComparison} shows that our work has a significant improvement in both latency and throughput compared with the previous best reported results for FPGAs.
The latency reported is for the entire system.

Remarkably, this is achieved with better accuracy compared to previous work as much higher precision activations are used.
Despite the low frequency, the current design already meets the performance of all existing comparable implementations even after normalizing for the large FPGA used in this paper.

It should be noted that for a fully floating point network, 37.3\,TOps/sec would be required to achieve the same number of classifications.
Only a fraction of these need to be executed in practice due to our compile-time optimisations as shown in the right two columns of Table~\ref{table:ops}.
This shows the reduction is mostly due to the implementation taking advantage of unstructured sparsity of 75\%.
This is further reduced by the application of CSE to the adder tree to reduce the remaining operations by removing the multiplications and merging computations.
Not reflected in the TOps but critical for the design is the 4 bit word serial adders and the bit serial adders which reduce area by factors of approximately 4 for layers 3 and 4 and 16 for layers 5 and 6, making this method feasible for the AWS F1 platform.

\subsection{Comparison with previous work}
The method proposed by Prost-Boucle et al. \cite{prost2017scalable} achieves the previously best reported low precision throughput on an FPGA for CIFAR10. 
This work implements a VGG-7 style network with ternary weights and activations.
The approach is similar to Baskin et al. \cite{baskin2018streaming} and Li et al. \cite{li20177} at a higher frequency of 250\,MHz and using hand written register transfer language (RTL) as opposed to C-based high-level synthesis (HLS) tools.
Their work is also the most similar to ours both in architecture and network choice, and they used an FPGA approximately $4\times$ smaller than ours. 
Compared with their work, our design achieves significantly higher FPS (122k FPS vs 27k FPS).
Accounting for the much larger FPGA used in this paper, this improvement in throughput disappears as the frequency is $2\times$ higher than achieved with this design.
The reason for the comparable throughput despite the advantage of exploiting sparsity is the use of 16-bit activations in this paper as opposed to ternary activations in Prost-Boucle et al.~\cite{prost2017scalable} and binary activations in Fraser et al.~\cite{fraser2017scaling}.
This design choice results in higher accuracy of 90.9\% on CIFAR10 compared to Prost-Boucle et al. with 86.7\% or Fraser et al. with 88.7\%.

The custom ASIC accelerator by Jouppi et al. \cite{jouppi2017datacenter}, was designed with datacenter requirements in mind.
Their work uses higher precision weights, meaning higher accuracy can probably be achieved.
Due to the much higher frequency of 700\,MHz compared to 125\,MHz and the difficulty of comparing the amount of hardware used it is hard to make a meaningful comparison on throughput.
Given the die area they report of 331\,mm$^2$ and the Ultrascale of 2256\,mm$^2$ the ASIC is significantly smaller.
Hence, in terms of throughput alone, the TPU is superior to this work.
However, the latency achieved by this design is very low.
While the TPU does not have a benchmark for CIFAR10, they quote a latency of around 10\,ms to achieve the peak throughput with a batch size of 250 images.
This is nearly 3 orders of magnitude more than the implementation in this paper.
For latency sensitive applications, this is a significant difference.
Additionally, the TPU is not commercially available for purchase and is accessible only through Google cloud.

YodaNN \cite{andri2017yodann} creates a very small ASIC with an area of 1.9\,mm$^2$.
They use binary weights and 12 bit activations and hence achieve a very high throughput for the area they used.
This is lower precision than the TPU and hence would likely lose some accuracy.
However, the throughput they get for the area is nearly $3\times$ that of the TPU and using 65~nm technology more than double the TPUs 28\,nm.

Venkatesh et. al. \cite{venkatesh2017accelerating} use 14nm technology to claim a peak speed of 2.5\,TOps/sec with a size of 1.09\,mm$^2$.
This uses ternary activations and half precision floating point, meaning it will likely have higher accuracy than YodaNN.

To summarise, the method proposed in this work achieves the highest throughput and the lowest latency reported on an FPGA platform so far, significantly reducing the gap to the highly customized ASIC architectures.

\begin{table*}[t]
\centering
\caption{ Comparison of CNN inference implementations for CIFAR10 where reported for ASICs (top) and FPGAs (bottom).}
\label{table:hardwareComparison}
\begin{tabular}{l l l r r r r l}
\toprule
Reference & Hardware & Precision & Freq. & Latency & TOps/sec & FPS & Accuracy\\
 & ($mm^2$,nm,LE$^5$/LC$^5$ $\times10^6$) & (wghts, actv) & [MHz] & & A/L/E$^6$ & & \\
\midrule
\cite{venkatesh2017accelerating} & ASIC(1.09,14,--) & (2,16$^2$) & 500 & -- & 2.5/2.5/2.5 & -- & 91.6\%$^3$ \\
\cite{andri2017yodann} & ASIC(1.9,65,--) & (1,12) & 480 & -- & 1.5/1.5/1.5 & 434 & -- \\
\cite{jouppi2017datacenter} & ASIC(331,28,--) & (8,8) & 700 & $\approx$10~ms & 86/86/86$^4$ & -- & -- \\
\cmidrule(r){1-8}
\cite{baskin2018streaming} & 5SGSD8(1600,28,0.7) &  (1,2) &105 & -- & -- & 1.2~k$^3$ & 84.2\% \\
\cite{li20177} & XC7VX690(1806.25,28,0.7) & (1$^1$, 1) & 90 & -- & 7.7/3.9/7.7 & 6.2~k  & 87.8\% \\
\cite{liang2018fp} & 5SGSD8(1600,28,0.7) & (1,1) & 150 & -- & 9.4/4.7/9.4 & ~7.6~k$^3$ & 86.31\% \\
\cite{prost2017scalable} & VC709(1806.25,28,0.7) & (2,2) & 250 & -- & 8.4/4.2/8.4 & 27~k & 86.7\% \\
\cite{umuroglu2017finn} & ZC706(961,28,0.35) & (1,1) & 200 & 283~\textmu s & 2.4/1.2/2.4 & 21.9~k & 80.1\%  \\
\cite{fraser2017scaling} & KU115(1600,20,1.45) & (1,1) & 125 & 671 \textmu s & 14.8/7.4/14.8 & 12~k & 88.7\% \\
This work & VU9P(2256.25,20,2.6) & (2,16) & 125 & 29~\textmu s & 2.5/2.5/37.3 & 122k & 90.9\% \\ 
\bottomrule
\end{tabular}
\caption*{$^1$First layer is fixed point,  $^2$floating point, $^3$estimated, $^4$ 92 TOps/sec peak, $^5$ LE and LC are from Xilinx or Altera documentation of the FPGAs, $^6$ Actual/Logical/Equivalent }
\end{table*}

\begin{figure}
    \centering
    \includegraphics[scale=0.5]{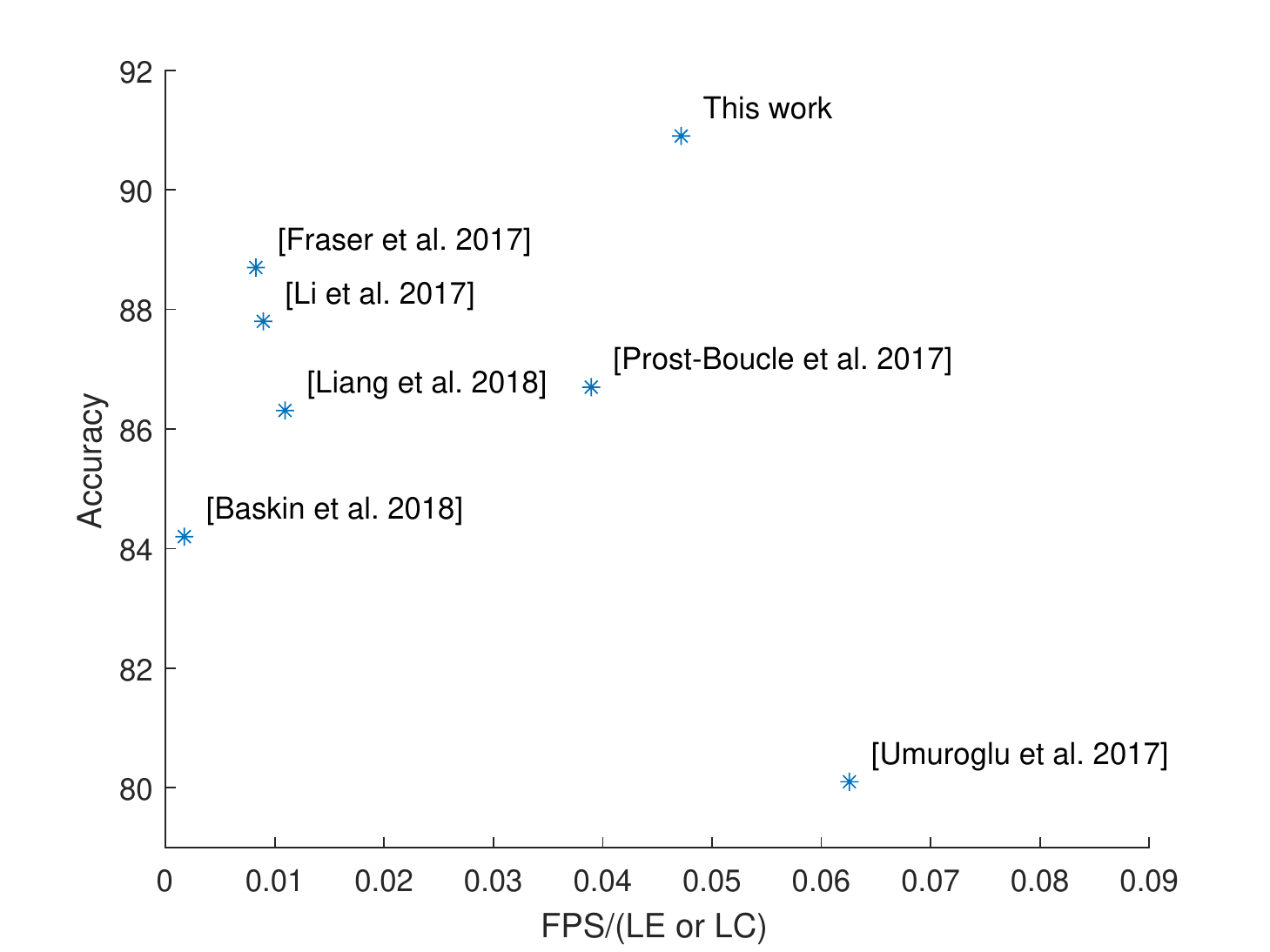}
    \caption{Comparison of FPGAs on CIFAR10}
    \label{fig:accr_throughput}
\end{figure}

Figure \ref{fig:accr_throughput} shows a comparison of accuracy and throughput normalized for Logic Elements (LE) and Logic Cells (LC).
This figure compares the available FPGA implementations on the CIFAR10 dataset.
Due to the higher precision activations chosen in this paper, the accuracy achieved by this design is significantly higher.
Despite the higher precision used, this design keeps up in throughput per LE or LC with other much lower precision implementations.


\section{Conclusion}
The method described of unrolling a convolution with ternary weights is very efficient for inference due to its ability to exploit unstructured sparsity.
Combined with the use of word or bit serial adders, this allows a very high image throughput and low latency.
While it is difficult to take advantage of sparsity on traditional computational platforms such as CPUs and GPUs, our method has no overhead for exploiting sparsity as redundancies are removed at compile time.
This paper has also demonstrated that the technique of Li et al. can directly tradeoff sparsity and accuracy.
Finally, as our architecture does not require the image to be buffered, larger images such as in ImageNet could still be used with the main constraints in implementation being the CNN size and the sparsity that can be achieved with acceptable accuracy.

While our technique is only loosely dependent on image size as there is only a limited amount of buffering, it is strongly dependent on the CNN size.
This technique has the disadvantage that compared to other approaches that can support larger networks this method is restricted by the amount of hardware available.
However, it has the advantage of very efficiently exploiting unstructured sparsity and common subexpressions.
Additionally, we do not require large batch sizes, resulting in a very low latency.
Our approach of unrolling the entire inference computation has only recently become feasible due to the increased size of FPGAs, improved quality of tools and research into low precision networks.
As FPGA capacity continues to increase, our method may become more favourable, particularly for small to medium neural networks.
Future work will involve creating an implementation of ImageNet which utilises multiple FPGAs, in particular using Amazon f1.16xlarge instances which contain 8 FPGAs.
Research into improving the merging of subexpressions for favourable routing in the convolutional layers should also be explored.

\section*{Acknowledgement} This work was partially funded by the Australia--Germany Joint Research Co--operation Scheme and the German Academic Exchange Service (DAAD) under grant no. 57388068.
This work was also supported by a CMCRC scholarship.

\appendix

\section{Supplementary Materials}
\label{sect:supp_materials}
The github repo at \url{https://github.com/da-steve101/binary_connect_cifar} is written in chisel3 that generates the VGG network in Verilog and Python scripts are used to verify the results and perform the CSE.
This generated code is used in \url{https://github.com/da-steve101/aws-fpga} and has an interface modified from the AWS cl\_sde example in branch sde\_if\_test.


\bibliographystyle{ACM-Reference-Format}
\bibliography{sample-bibliography}

%% file: sample-journal.bbl

\begin{thebibliography}{34}


\ifx \showCODEN    \undefined \def \showCODEN     #1{\unskip}     \fi
\ifx \showDOI      \undefined \def \showDOI       #1{#1}\fi
\ifx \showISBNx    \undefined \def \showISBNx     #1{\unskip}     \fi
\ifx \showISBNxiii \undefined \def \showISBNxiii  #1{\unskip}     \fi
\ifx \showISSN     \undefined \def \showISSN      #1{\unskip}     \fi
\ifx \showLCCN     \undefined \def \showLCCN      #1{\unskip}     \fi
\ifx \shownote     \undefined \def \shownote      #1{#1}          \fi
\ifx \showarticletitle \undefined \def \showarticletitle #1{#1}   \fi
\ifx \showURL      \undefined \def \showURL       {\relax}        \fi
\providecommand\bibfield[2]{#2}
\providecommand\bibinfo[2]{#2}
\providecommand\natexlab[1]{#1}
\providecommand\showeprint[2][]{arXiv:#2}

\bibitem[\protect\citeauthoryear{Andri, Cavigelli, Rossi, and Benini}{Andri
  et~al\mbox{.}}{2017}]%
        {andri2017yodann}
\bibfield{author}{\bibinfo{person}{Renzo Andri}, \bibinfo{person}{Lukas
  Cavigelli}, \bibinfo{person}{Davide Rossi}, {and} \bibinfo{person}{Luca
  Benini}.} \bibinfo{year}{2017}\natexlab{}.
\newblock \showarticletitle{YodaNN: An architecture for ultralow power
  binary-weight CNN acceleration}.
\newblock \bibinfo{journal}{\emph{IEEE Transactions on Computer-Aided Design of
  Integrated Circuits and Systems}} \bibinfo{volume}{37}, \bibinfo{number}{1}
  (\bibinfo{year}{2017}), \bibinfo{pages}{48--60}.
\newblock


\bibitem[\protect\citeauthoryear{Bachrach, Vo, Richards, Lee, Waterman,
  Avi{\v{z}}ienis, Wawrzynek, and Asanovi{\'c}}{Bachrach et~al\mbox{.}}{2012}]%
        {bachrach2012chisel}
\bibfield{author}{\bibinfo{person}{Jonathan Bachrach}, \bibinfo{person}{Huy
  Vo}, \bibinfo{person}{Brian Richards}, \bibinfo{person}{Yunsup Lee},
  \bibinfo{person}{Andrew Waterman}, \bibinfo{person}{Rimas Avi{\v{z}}ienis},
  \bibinfo{person}{John Wawrzynek}, {and} \bibinfo{person}{Krste
  Asanovi{\'c}}.} \bibinfo{year}{2012}\natexlab{}.
\newblock \showarticletitle{Chisel: constructing hardware in a scala embedded
  language}. In \bibinfo{booktitle}{\emph{DAC Design Automation Conference
  2012}}. IEEE, \bibinfo{pages}{1212--1221}.
\newblock


\bibitem[\protect\citeauthoryear{Baskin, Liss, Zheltonozhskii, Bronstein, and
  Mendelson}{Baskin et~al\mbox{.}}{2018}]%
        {baskin2018streaming}
\bibfield{author}{\bibinfo{person}{Chaim Baskin}, \bibinfo{person}{Natan Liss},
  \bibinfo{person}{Evgenii Zheltonozhskii}, \bibinfo{person}{Alex~M Bronstein},
  {and} \bibinfo{person}{Avi Mendelson}.} \bibinfo{year}{2018}\natexlab{}.
\newblock \showarticletitle{Streaming architecture for large-scale quantized
  neural networks on an FPGA-based dataflow platform}. In
  \bibinfo{booktitle}{\emph{2018 IEEE International Parallel and Distributed
  Processing Symposium Workshops (IPDPSW)}}. IEEE, \bibinfo{pages}{162--169}.
\newblock


\bibitem[\protect\citeauthoryear{Boo and Sung}{Boo and Sung}{2017}]%
        {boo2017structured}
\bibfield{author}{\bibinfo{person}{Yoonho Boo} {and} \bibinfo{person}{Wonyong
  Sung}.} \bibinfo{year}{2017}\natexlab{}.
\newblock \showarticletitle{Structured sparse ternary weight coding of deep
  neural networks for efficient hardware implementations}. In
  \bibinfo{booktitle}{\emph{2017 IEEE International workshop on signal
  processing systems (SiPS)}}. IEEE, \bibinfo{pages}{1--6}.
\newblock


\bibitem[\protect\citeauthoryear{Cappello and Steiglitz}{Cappello and
  Steiglitz}{1984}]%
        {cs84}
\bibfield{author}{\bibinfo{person}{P Cappello} {and} \bibinfo{person}{K
  Steiglitz}.} \bibinfo{year}{1984}\natexlab{}.
\newblock \showarticletitle{{Some Complexity Issues in Digital Signal
  Processing}}.
\newblock \bibinfo{journal}{\emph{IEEE Transactions on Acoustics, Speech and
  Signal Processing}} \bibinfo{volume}{32}, \bibinfo{number}{5}
  (\bibinfo{date}{Oct.} \bibinfo{year}{1984}), \bibinfo{pages}{1037--1041}.
\newblock


\bibitem[\protect\citeauthoryear{Chen, Krishna, Emer, and Sze}{Chen
  et~al\mbox{.}}{2017}]%
        {chen2017eyeriss}
\bibfield{author}{\bibinfo{person}{Yu-Hsin Chen}, \bibinfo{person}{Tushar
  Krishna}, \bibinfo{person}{Joel~S Emer}, {and} \bibinfo{person}{Vivienne
  Sze}.} \bibinfo{year}{2017}\natexlab{}.
\newblock \showarticletitle{Eyeriss: An energy-efficient reconfigurable
  accelerator for deep convolutional neural networks}.
\newblock \bibinfo{journal}{\emph{IEEE Journal of Solid-State Circuits}}
  \bibinfo{volume}{52}, \bibinfo{number}{1} (\bibinfo{year}{2017}),
  \bibinfo{pages}{127--138}.
\newblock


\bibitem[\protect\citeauthoryear{Courbariaux, Hubara, Soudry, El-Yaniv, and
  Bengio}{Courbariaux et~al\mbox{.}}{2016}]%
        {courbariaux2016binarized}
\bibfield{author}{\bibinfo{person}{Matthieu Courbariaux}, \bibinfo{person}{Itay
  Hubara}, \bibinfo{person}{Daniel Soudry}, \bibinfo{person}{Ran El-Yaniv},
  {and} \bibinfo{person}{Yoshua Bengio}.} \bibinfo{year}{2016}\natexlab{}.
\newblock \showarticletitle{Binarized neural networks: Training deep neural
  networks with weights and activations constrained to+ 1 or-1}.
\newblock \bibinfo{journal}{\emph{arXiv preprint arXiv:1602.02830}}
  (\bibinfo{year}{2016}).
\newblock


\bibitem[\protect\citeauthoryear{Faraone, Fraser, Gambardella, Blott, and
  Leong}{Faraone et~al\mbox{.}}{2017}]%
        {faraone2017compressing}
\bibfield{author}{\bibinfo{person}{Julian Faraone}, \bibinfo{person}{Nicholas
  Fraser}, \bibinfo{person}{Giulio Gambardella}, \bibinfo{person}{Michaela
  Blott}, {and} \bibinfo{person}{Philip~HW Leong}.}
  \bibinfo{year}{2017}\natexlab{}.
\newblock \showarticletitle{Compressing Low Precision Deep Neural Networks
  Using Sparsity-Induced Regularization in Ternary Networks}. In
  \bibinfo{booktitle}{\emph{International Conference on Neural Information
  Processing}}. Springer, \bibinfo{pages}{393--404}.
\newblock


\bibitem[\protect\citeauthoryear{Foundation}{Foundation}{2015}]%
        {dpdk}
\bibfield{author}{\bibinfo{person}{Linux Foundation}.}
  \bibinfo{year}{2015}\natexlab{}.
\newblock \bibinfo{title}{Data Plane Development Kit ({DPDK})}.
\newblock   (\bibinfo{year}{2015}).
\newblock
\urldef\tempurl%
\url{http://www.dpdk.org}
\showURL{%
\tempurl}


\bibitem[\protect\citeauthoryear{Fraser, Umuroglu, Gambardella, Blott, Leong,
  Jahre, and Vissers}{Fraser et~al\mbox{.}}{2017}]%
        {fraser2017scaling}
\bibfield{author}{\bibinfo{person}{Nicholas~J Fraser}, \bibinfo{person}{Yaman
  Umuroglu}, \bibinfo{person}{Giulio Gambardella}, \bibinfo{person}{Michaela
  Blott}, \bibinfo{person}{Philip Leong}, \bibinfo{person}{Magnus Jahre}, {and}
  \bibinfo{person}{Kees Vissers}.} \bibinfo{year}{2017}\natexlab{}.
\newblock \showarticletitle{Scaling binarized neural networks on reconfigurable
  logic}. In \bibinfo{booktitle}{\emph{Proc. 8th Workshop and 6th Workshop on
  Parallel Programming and Run-Time Management Techniques for Many-core
  Architectures and Design Tools and Architectures for Multicore Embedded
  Computing Platforms}}. ACM, \bibinfo{pages}{25--30}.
\newblock


\bibitem[\protect\citeauthoryear{Graham}{Graham}{2014}]%
        {graham2014fractional}
\bibfield{author}{\bibinfo{person}{Ben Graham}.}
  \bibinfo{year}{2014}\natexlab{}.
\newblock \showarticletitle{Fractional max-pooling (2014)}.
\newblock \bibinfo{journal}{\emph{arXiv preprint arXiv:1412.6071}}
  (\bibinfo{year}{2014}).
\newblock


\bibitem[\protect\citeauthoryear{Hardieck, Kumm, M{\"o}ller, and Zipf}{Hardieck
  et~al\mbox{.}}{2018}]%
        {Hardieck2018}
\bibfield{author}{\bibinfo{person}{Martin Hardieck}, \bibinfo{person}{Martin
  Kumm}, \bibinfo{person}{Konrad M{\"o}ller}, {and} \bibinfo{person}{Peter
  Zipf}.} \bibinfo{year}{2018}\natexlab{}.
\newblock \showarticletitle{{\emph{Constant Matrix Multiplication with Ternary
  Adders}}}. In \bibinfo{booktitle}{\emph{IEEE International Conference on
  Electronics, Circuits and Systems (ICECS)}}.
\newblock


\bibitem[\protect\citeauthoryear{Hsiao, Chen, and Tu}{Hsiao
  et~al\mbox{.}}{2006}]%
        {hsiao2006memory}
\bibfield{author}{\bibinfo{person}{Shen-Fu Hsiao}, \bibinfo{person}{Ming-Chih
  Chen}, {and} \bibinfo{person}{Chia-Shin Tu}.}
  \bibinfo{year}{2006}\natexlab{}.
\newblock \showarticletitle{Memory-free low-cost designs of advanced encryption
  standard using common subexpression elimination for subfunctions in
  transformations}.
\newblock \bibinfo{journal}{\emph{IEEE Transactions on Circuits and Systems I:
  Regular Papers}} \bibinfo{volume}{53}, \bibinfo{number}{3}
  (\bibinfo{year}{2006}), \bibinfo{pages}{615--626}.
\newblock


\bibitem[\protect\citeauthoryear{Jouppi, Young, Patil, Patterson, Agrawal,
  Bajwa, Bates, Bhatia, Boden, Borchers, et~al\mbox{.}}{Jouppi
  et~al\mbox{.}}{2017}]%
        {jouppi2017datacenter}
\bibfield{author}{\bibinfo{person}{Norman~P Jouppi}, \bibinfo{person}{Cliff
  Young}, \bibinfo{person}{Nishant Patil}, \bibinfo{person}{David Patterson},
  \bibinfo{person}{Gaurav Agrawal}, \bibinfo{person}{Raminder Bajwa},
  \bibinfo{person}{Sarah Bates}, \bibinfo{person}{Suresh Bhatia},
  \bibinfo{person}{Nan Boden}, \bibinfo{person}{Al Borchers}, {et~al\mbox{.}}}
  \bibinfo{year}{2017}\natexlab{}.
\newblock \showarticletitle{In-datacenter performance analysis of a tensor
  processing unit}. In \bibinfo{booktitle}{\emph{Proc. 44th Annual
  International Symposium on Computer Architecture}}. ACM,
  \bibinfo{pages}{1--12}.
\newblock


\bibitem[\protect\citeauthoryear{Kim, Grady, Lian, Brothers, and Anderson}{Kim
  et~al\mbox{.}}{2017}]%
        {kimfpga}
\bibfield{author}{\bibinfo{person}{Jin~Hee Kim}, \bibinfo{person}{Brett Grady},
  \bibinfo{person}{Ruolong Lian}, \bibinfo{person}{John Brothers}, {and}
  \bibinfo{person}{Jason~H Anderson}.} \bibinfo{year}{2017}\natexlab{}.
\newblock \showarticletitle{{FPGA}-Based {CNN} Inference Accelerator
  Synthesized from Multi-Threaded {C} Software}.
\newblock  (\bibinfo{year}{2017}).
\newblock


\bibitem[\protect\citeauthoryear{Krizhevsky and Hinton}{Krizhevsky and
  Hinton}{2009}]%
        {krizhevsky2009learning}
\bibfield{author}{\bibinfo{person}{Alex Krizhevsky} {and}
  \bibinfo{person}{Geoffrey Hinton}.} \bibinfo{year}{2009}\natexlab{}.
\newblock \bibinfo{booktitle}{\emph{Learning multiple layers of features from
  tiny images}}.
\newblock \bibinfo{type}{{T}echnical {R}eport}.
  \bibinfo{institution}{Citeseer}.
\newblock


\bibitem[\protect\citeauthoryear{Kumm, Hardieck, and Zipf}{Kumm
  et~al\mbox{.}}{2017}]%
        {kumm2017}
\bibfield{author}{\bibinfo{person}{Martin Kumm}, \bibinfo{person}{Martin
  Hardieck}, {and} \bibinfo{person}{Peter Zipf}.}
  \bibinfo{year}{2017}\natexlab{}.
\newblock \showarticletitle{{Optimization of Constant Matrix Multiplication
  with Low Power and High Throughput}}.
\newblock \bibinfo{journal}{\emph{IEEE Trans. Comput.}} \bibinfo{volume}{66},
  \bibinfo{number}{12} (\bibinfo{year}{2017}), \bibinfo{pages}{2072--2080}.
\newblock


\bibitem[\protect\citeauthoryear{Kumm, Zipf, Faust, and Chang}{Kumm
  et~al\mbox{.}}{2012}]%
        {kumm2012pipelined}
\bibfield{author}{\bibinfo{person}{Martin Kumm}, \bibinfo{person}{Peter Zipf},
  \bibinfo{person}{Mathias Faust}, {and} \bibinfo{person}{Chip-Hong Chang}.}
  \bibinfo{year}{2012}\natexlab{}.
\newblock \showarticletitle{Pipelined adder graph optimization for high speed
  multiple constant multiplication}. In \bibinfo{booktitle}{\emph{Circuits and
  Systems (ISCAS), 2012 IEEE International Symposium on}}. IEEE,
  \bibinfo{pages}{49--52}.
\newblock


\bibitem[\protect\citeauthoryear{LeCun, Bengio, and Hinton}{LeCun
  et~al\mbox{.}}{2015}]%
        {DL}
\bibfield{author}{\bibinfo{person}{Yann LeCun}, \bibinfo{person}{Yoshua
  Bengio}, {and} \bibinfo{person}{Geoffrey Hinton}.}
  \bibinfo{year}{2015}\natexlab{}.
\newblock \showarticletitle{Deep learning}.
\newblock \bibinfo{journal}{\emph{Nature}} \bibinfo{volume}{521},
  \bibinfo{number}{7553} (\bibinfo{date}{5} \bibinfo{year}{2015}),
  \bibinfo{pages}{436--444}.
\newblock
\showISSN{0028-0836}
\urldef\tempurl%
\url{https://doi.org/10.1038/nature14539}
\showDOI{\tempurl}


\bibitem[\protect\citeauthoryear{Li, Zhang, and Liu}{Li et~al\mbox{.}}{2016}]%
        {li2016ternary}
\bibfield{author}{\bibinfo{person}{Fengfu Li}, \bibinfo{person}{Bo Zhang},
  {and} \bibinfo{person}{Bin Liu}.} \bibinfo{year}{2016}\natexlab{}.
\newblock \showarticletitle{Ternary weight networks}.
\newblock \bibinfo{journal}{\emph{arXiv preprint arXiv:1605.04711}}
  (\bibinfo{year}{2016}).
\newblock


\bibitem[\protect\citeauthoryear{Li, Liu, Xu, Yu, and Ren}{Li
  et~al\mbox{.}}{2017}]%
        {li20177}
\bibfield{author}{\bibinfo{person}{Yixing Li}, \bibinfo{person}{Zichuan Liu},
  \bibinfo{person}{Kai Xu}, \bibinfo{person}{Hao Yu}, {and}
  \bibinfo{person}{Fengbo Ren}.} \bibinfo{year}{2017}\natexlab{}.
\newblock \showarticletitle{A 7.663-TOPS 8.2-W energy-efficient FPGA
  accelerator for binary convolutional neural networks}. In
  \bibinfo{booktitle}{\emph{FPGA}}. \bibinfo{pages}{290--291}.
\newblock


\bibitem[\protect\citeauthoryear{Liang, Yin, Liu, Luk, and Wei}{Liang
  et~al\mbox{.}}{2018}]%
        {liang2018fp}
\bibfield{author}{\bibinfo{person}{Shuang Liang}, \bibinfo{person}{Shouyi Yin},
  \bibinfo{person}{Leibo Liu}, \bibinfo{person}{Wayne Luk}, {and}
  \bibinfo{person}{Shaojun Wei}.} \bibinfo{year}{2018}\natexlab{}.
\newblock \showarticletitle{FP-BNN: Binarized neural network on FPGA}.
\newblock \bibinfo{journal}{\emph{Neurocomputing}}  \bibinfo{volume}{275}
  (\bibinfo{year}{2018}), \bibinfo{pages}{1072--1086}.
\newblock


\bibitem[\protect\citeauthoryear{Mellempudi, Kundu, Mudigere, Das, Kaul, and
  Dubey}{Mellempudi et~al\mbox{.}}{2017}]%
        {mellempudi2017ternary}
\bibfield{author}{\bibinfo{person}{Naveen Mellempudi}, \bibinfo{person}{Abhisek
  Kundu}, \bibinfo{person}{Dheevatsa Mudigere}, \bibinfo{person}{Dipankar Das},
  \bibinfo{person}{Bharat Kaul}, {and} \bibinfo{person}{Pradeep Dubey}.}
  \bibinfo{year}{2017}\natexlab{}.
\newblock \showarticletitle{Ternary Neural Networks with Fine-Grained
  Quantization}.
\newblock \bibinfo{journal}{\emph{arXiv preprint arXiv:1705.01462}}
  (\bibinfo{year}{2017}).
\newblock


\bibitem[\protect\citeauthoryear{Meloni, Capotondi, Deriu, Brian, Conti, Rossi,
  Raffo, and Benini}{Meloni et~al\mbox{.}}{2018}]%
        {meloni2018neura}
\bibfield{author}{\bibinfo{person}{Paolo Meloni}, \bibinfo{person}{Alessandro
  Capotondi}, \bibinfo{person}{Gianfranco Deriu}, \bibinfo{person}{Michele
  Brian}, \bibinfo{person}{Francesco Conti}, \bibinfo{person}{Davide Rossi},
  \bibinfo{person}{Luigi Raffo}, {and} \bibinfo{person}{Luca Benini}.}
  \bibinfo{year}{2018}\natexlab{}.
\newblock \showarticletitle{NEURA ghe: Exploiting CPU-FPGA Synergies for
  Efficient and Flexible CNN Inference Acceleration on Zynq SoCs}.
\newblock \bibinfo{journal}{\emph{ACM Transactions on Reconfigurable Technology
  and Systems (TRETS)}} \bibinfo{volume}{11}, \bibinfo{number}{3}
  (\bibinfo{year}{2018}), \bibinfo{pages}{18}.
\newblock


\bibitem[\protect\citeauthoryear{Moss, Nurvitadhi, Sim, Mishra, Marr,
  Subhaschandra, and Leong}{Moss et~al\mbox{.}}{2017}]%
        {moss2017high}
\bibfield{author}{\bibinfo{person}{Duncan~JM Moss}, \bibinfo{person}{Eriko
  Nurvitadhi}, \bibinfo{person}{Jaewoong Sim}, \bibinfo{person}{Asit Mishra},
  \bibinfo{person}{Debbie Marr}, \bibinfo{person}{Suchit Subhaschandra}, {and}
  \bibinfo{person}{Philip~HW Leong}.} \bibinfo{year}{2017}\natexlab{}.
\newblock \showarticletitle{High performance binary neural networks on the
  {Xeon+ FPGA}{\texttrademark} platform}. In \bibinfo{booktitle}{\emph{Field
  Programmable Logic and Applications (FPL), 2017 27th International Conference
  on}}. IEEE, \bibinfo{pages}{1--4}.
\newblock


\bibitem[\protect\citeauthoryear{Prost-Boucle, Bourge, P{\'e}trot, Alemdar,
  Caldwell, and Leroy}{Prost-Boucle et~al\mbox{.}}{2017}]%
        {prost2017scalable}
\bibfield{author}{\bibinfo{person}{Adrien Prost-Boucle}, \bibinfo{person}{Alban
  Bourge}, \bibinfo{person}{Fr{\'e}d{\'e}ric P{\'e}trot},
  \bibinfo{person}{Hande Alemdar}, \bibinfo{person}{Nicholas Caldwell}, {and}
  \bibinfo{person}{Vincent Leroy}.} \bibinfo{year}{2017}\natexlab{}.
\newblock \showarticletitle{Scalable high-performance architecture for
  convolutional ternary neural networks on {FPGA}}. In
  \bibinfo{booktitle}{\emph{Field Programmable Logic and Applications (FPL),
  2017 27th International Conference on}}. IEEE, \bibinfo{pages}{1--7}.
\newblock


\bibitem[\protect\citeauthoryear{Qiu, Wang, Yao, Guo, Li, Zhou, Yu, Tang, Xu,
  Song, et~al\mbox{.}}{Qiu et~al\mbox{.}}{2016}]%
        {qiu2016going}
\bibfield{author}{\bibinfo{person}{Jiantao Qiu}, \bibinfo{person}{Jie Wang},
  \bibinfo{person}{Song Yao}, \bibinfo{person}{Kaiyuan Guo},
  \bibinfo{person}{Boxun Li}, \bibinfo{person}{Erjin Zhou},
  \bibinfo{person}{Jincheng Yu}, \bibinfo{person}{Tianqi Tang},
  \bibinfo{person}{Ningyi Xu}, \bibinfo{person}{Sen Song}, {et~al\mbox{.}}}
  \bibinfo{year}{2016}\natexlab{}.
\newblock \showarticletitle{Going deeper with embedded fpga platform for
  convolutional neural network}. In \bibinfo{booktitle}{\emph{Proc. 2016
  ACM/SIGDA International Symposium on Field-Programmable Gate Arrays}}. ACM,
  \bibinfo{pages}{26--35}.
\newblock


\bibitem[\protect\citeauthoryear{Rastegari, Ordonez, Redmon, and
  Farhadi}{Rastegari et~al\mbox{.}}{2016}]%
        {rastegari2016xnor}
\bibfield{author}{\bibinfo{person}{Mohammad Rastegari},
  \bibinfo{person}{Vicente Ordonez}, \bibinfo{person}{Joseph Redmon}, {and}
  \bibinfo{person}{Ali Farhadi}.} \bibinfo{year}{2016}\natexlab{}.
\newblock \showarticletitle{Xnor-net: Imagenet classification using binary
  convolutional neural networks}. In \bibinfo{booktitle}{\emph{European
  Conference on Computer Vision}}. Springer, \bibinfo{pages}{525--542}.
\newblock


\bibitem[\protect\citeauthoryear{Umuroglu, Fraser, Gambardella, Blott, Leong,
  Jahre, and Vissers}{Umuroglu et~al\mbox{.}}{2017}]%
        {umuroglu2017finn}
\bibfield{author}{\bibinfo{person}{Yaman Umuroglu}, \bibinfo{person}{Nicholas~J
  Fraser}, \bibinfo{person}{Giulio Gambardella}, \bibinfo{person}{Michaela
  Blott}, \bibinfo{person}{Philip Leong}, \bibinfo{person}{Magnus Jahre}, {and}
  \bibinfo{person}{Kees Vissers}.} \bibinfo{year}{2017}\natexlab{}.
\newblock \showarticletitle{Finn: A framework for fast, scalable binarized
  neural network inference}. In \bibinfo{booktitle}{\emph{Proc. 2017 ACM/SIGDA
  International Symposium on Field-Programmable Gate Arrays}}. ACM,
  \bibinfo{pages}{65--74}.
\newblock


\bibitem[\protect\citeauthoryear{Venkatesh, Nurvitadhi, and Marr}{Venkatesh
  et~al\mbox{.}}{2017}]%
        {venkatesh2017accelerating}
\bibfield{author}{\bibinfo{person}{Ganesh Venkatesh}, \bibinfo{person}{Eriko
  Nurvitadhi}, {and} \bibinfo{person}{Debbie Marr}.}
  \bibinfo{year}{2017}\natexlab{}.
\newblock \showarticletitle{Accelerating deep convolutional networks using
  low-precision and sparsity}. In \bibinfo{booktitle}{\emph{Acoustics, Speech
  and Signal Processing (ICASSP), 2017 IEEE International Conference on}}.
  IEEE, \bibinfo{pages}{2861--2865}.
\newblock


\bibitem[\protect\citeauthoryear{Wang, Liang, Yao, Shan, Han, Peng, and
  Luo}{Wang et~al\mbox{.}}{2017}]%
        {Wang2017ReconfigurablePF}
\bibfield{author}{\bibinfo{person}{Yanshu Wang}, \bibinfo{person}{Shuang
  Liang}, \bibinfo{person}{Song Yao}, \bibinfo{person}{Yi Shan},
  \bibinfo{person}{Song Han}, \bibinfo{person}{Junjie Peng}, {and}
  \bibinfo{person}{Hong~Xia Luo}.} \bibinfo{year}{2017}\natexlab{}.
\newblock \showarticletitle{Reconfigurable Processor for Deep Learning in
  Autonomous Vehicles}.
\newblock


\bibitem[\protect\citeauthoryear{Wu, Zhang, Ye, and Lan}{Wu
  et~al\mbox{.}}{2013}]%
        {wu2013improving}
\bibfield{author}{\bibinfo{person}{Ning Wu}, \bibinfo{person}{Xiaoqiang Zhang},
  \bibinfo{person}{Yunfei Ye}, {and} \bibinfo{person}{Lidong Lan}.}
  \bibinfo{year}{2013}\natexlab{}.
\newblock \showarticletitle{Improving Common Subexpression Elimination
  Algorithm with A New Gate-Level Delay Computing Method}. In
  \bibinfo{booktitle}{\emph{Proceedings of the World Congress on Engineering
  and Computer Science}}, Vol.~\bibinfo{volume}{2}.
\newblock


\bibitem[\protect\citeauthoryear{Zhang and Prasanna}{Zhang and
  Prasanna}{2017}]%
        {zhang2017frequency}
\bibfield{author}{\bibinfo{person}{Chi Zhang} {and} \bibinfo{person}{Viktor
  Prasanna}.} \bibinfo{year}{2017}\natexlab{}.
\newblock \showarticletitle{Frequency domain acceleration of convolutional
  neural networks on {CPU-FPGA} shared memory system}. In
  \bibinfo{booktitle}{\emph{Proc. 2017 ACM/SIGDA International Symposium on
  Field-Programmable Gate Arrays}}. ACM, \bibinfo{pages}{35--44}.
\newblock


\bibitem[\protect\citeauthoryear{Zhou, Wu, Ni, Zhou, Wen, and Zou}{Zhou
  et~al\mbox{.}}{2016}]%
        {zhou2016dorefa}
\bibfield{author}{\bibinfo{person}{Shuchang Zhou}, \bibinfo{person}{Yuxin Wu},
  \bibinfo{person}{Zekun Ni}, \bibinfo{person}{Xinyu Zhou}, \bibinfo{person}{He
  Wen}, {and} \bibinfo{person}{Yuheng Zou}.} \bibinfo{year}{2016}\natexlab{}.
\newblock \showarticletitle{DoReFa-Net: Training low bitwidth convolutional
  neural networks with low bitwidth gradients}.
\newblock \bibinfo{journal}{\emph{arXiv preprint arXiv:1606.06160}}
  (\bibinfo{year}{2016}).
\newblock


\end{thebibliography}
